\documentclass[a4paper,fleqn,usenatbib]{mnras}
\usepackage{Times}
\usepackage[T1]{fontenc}
\usepackage{ae,aecompl}

\usepackage{graphicx}	
\usepackage{amsmath}	
\usepackage{amssymb}	

\title[Evolution of the UV LF and SMF of LAEs]{The evolution of the UV luminosity and stellar mass functions of Lyman-$\alpha$ emitters from $z\sim2$ to $z\sim6$}

\author[Santos et al.]{S. Santos$^{1}$\thanks{E-mail: s.santos@lancaster.ac.uk}, D. Sobral$^{1}$, J. Butterworth$^{1,3}$, A. Paulino-Afonso$^{2}$, B. Ribeiro$^{3}$,\newauthor E. da Cunha$^{4,5,6}$, J. Calhau$^{1,9,10}$, A. A. Khostovan$^{7}$\thanks{NASA Postdoctoral Program Fellow}, J. Matthee$^{8}$\thanks{Zwicky Fellow}, P. Arrabal Haro$^{9,10}$\\ 
$^{1}$ Department of Physics, Lancaster University, Lancaster, LA1 4YB, UK\\
$^{2}$ CENTRA - Centro de Astrof\'{i}sica e Gravita\c{c}\~{a}o, Instituto Superior T\'{e}cnico, Av. Rovisco Pais, 1, 1049-001, Lisboa, Portugal\\
$^{3}$ Leiden Observatory, Leiden University, P.O. Box 9513, NL-2300 RA Leiden, The Netherlands\\
$^{4}$ International Centre for Radio Astronomy Research, University of Western Australia, 35 Stirling Hwy, Crawley, WA 6009, Australia \\
$^{5}$ Research School of Astronomy and Astrophysics, The Australian National University, Canberra, ACT 2611, Australia \\
$^{6}$ ARC Centre of Excellence for All Sky Astrophysics in 3 Dimensions (ASTRO 3D), Millers Point NSW 2000, Australia\\
$^{7}$ Astrophysics Division, NASA Goddard Space Flight Center, Greenbelt, MD 20771, United States of America\\
$^{8}$ Department of Physics, ETH Z$\ddot{u}$rich,Wolfgang-Pauli-Strasse 27, 8093 Z$\ddot{u}$rich, Switzerland\\
$^{9}$ Instituto de Astrof\'{i}sica de Canarias (IAC), E-38200 La Laguna, Tenerife, Spain\\
$^{10}$ Departamento de Astrof\'{i}sica, Universidad de La Laguna, E-38206 La Laguna, Tenerife, Spain}

\pubyear{2019}

\begin{document}
\label{firstpage}
\pagerange{\pageref{firstpage}--\pageref{lastpage}}
\maketitle

\begin{abstract}
We measure the evolution of the rest-frame UV luminosity function (LF) and the stellar mass function (SMF) of Lyman-$\alpha$ (Ly$\alpha$) emitters (LAEs) from $z\sim2$ to $z\sim6$ by exploring $\sim4000$ LAEs from the SC4K sample. We find a correlation between Ly$\alpha$ luminosity (L$_{\rm Ly\alpha}$) and rest-frame UV (M$_{\rm UV}$), with best-fit M$_{\rm UV}=-1.6_{-0.3}^{+0.2}\log_{10} (\rm L_{Ly\alpha}/erg\,s^{-1})+47_{-11}^{+12}$ and a shallower relation between L$_{\rm Ly\alpha}$ and stellar mass (M$_\star$), with best-fit $\log_{10} (\rm M_\star/M_\odot)=0.9_{-0.1}^{+0.1}\log_{10} (\rm L_{Ly\alpha}/erg\,s^{-1})-28_{-3.8}^{+4.0}$. An increasing L$_{\rm Ly\alpha}$ cut predominantly lowers the number density of faint M$_{\rm UV}$ and low M$_\star$ LAEs. We estimate a proxy for the full UV LFs and SMFs of LAEs with simple assumptions of the faint end slope. For the UV LF, we find a brightening of the characteristic UV luminosity (M$_{\rm UV}^*$) with increasing redshift and a decrease of the characteristic number density ($\Phi^*$). For the SMF, we measure a characteristic stellar mass (${\rm M_\star^*/M_\odot}$) increase with increasing redshift, and a  $\Phi^*$ decline. However, if we apply a uniform luminosity cut of $\log_{10} (\rm L_{Ly\alpha}/erg\,s^{-1}) \geq 43.0$, we find much milder to no evolution in the UV and SMF of LAEs. The UV luminosity density ($\rho_{\rm UV}$) of the full sample of LAEs shows moderate evolution and the stellar mass density ($\rho_{\rm M}$) decreases, with both being always lower than the total $\rho_{\rm UV}$ and $\rho_{\rm M}$ of more typical galaxies but slowly approaching them with increasing redshift. Overall, our results indicate that both $\rho_{\rm UV}$ and $\rho_{\rm M}$ of LAEs slowly approach the measurements of continuum-selected galaxies at $z>6$, which suggests a key role of LAEs in the epoch of reionisation.
\end{abstract}


\section{Introduction} \label{sec:introduction}

Multiple studies have used the Lyman-$\alpha$ (Ly$\alpha$, $\lambda_{\rm 0,vacuum}=1215.67$\,\AA) emission line to successfully select large samples of galaxies at $z>2$ \citep[e.g.][]{Cowie1998,Malhotra2004,vanBreukelen2005,Ouchi2008,Rauch2008, Matthee2015, Santos2016,Drake2017, Sobral2018,Konno2018,Taylor2020}. Ly$\alpha$ emission is typically associated with young star-forming galaxies \citep[SFGs, e.g.][]{PartridgePeebles1967} but can also be emitted from active galaxy nuclei \citep[AGN; e.g.][]{Miley2008,Sobral2018Spectra,Calhau2020}.

LAEs are typically young/primeval, low mass, low dust extinction sources \citep[e.g.][]{Gawiser2006,Gawiser2007,Pentericci2007,Lai2008,Matthee2021}, but a significant diversity of properties within the Ly$\alpha$ population has been reported in the literature \citep[e.g.][]{Lai2008,Finkelstein2009,Acquaviva2012,Hagen2016,Matthee2016,Oyarzun2017,Santos2020}. Sources with high Ly$\alpha$ equivalent width (EW) typically have young stellar ages, low metallicities and top-heavy initial mass functions \citep[e.g.][]{Schaerer2003,Raiter2010,Hashimoto2017}. LAEs have been shown to be very compact in the UV \citep[e.g.][]{Malhotra2012,PaulinoAfonso2018}, with the compact morphology possibly being favourable to the escape of Ly$\alpha$ photons. Additionally, studies have shown that high-redshift LAEs may be progenitors of a wide range of galaxies, from present-day galaxies \citep[e.g.][]{Gawiser2007,Guaita2010,Yajima2012} to bright cluster galaxies \citep[BCGs; e.g.][]{Khostovan2019}, highlighting the significance of LAEs in galaxy evolution studies.

Studies of UV-continuum selected galaxies have found that the Ly$\alpha$ fraction ($\chi_{\rm Ly\alpha}$, percentage of galaxies with Ly$\alpha$ emission) increases with redshift up to $z\sim6$ \citep[e.g.][]{Stark2010,Pentericci2011,Cassata2015,deBarros2017,Caruana2018,Kusakabe2020}. This might be explained by an average lower dust content in higher redshift galaxies \citep[e.g.][]{Stanway2005,Bouwens2006}, increasing the Ly$\alpha$ escape fraction \citep[f$_{\rm esc, Ly\alpha}$, ratio between observed and intrinsic Ly$\alpha$ photons in a galaxy; e.g.][]{Hayes2011} and/or increasing the ionising efficiency \citep[$\xi_{\rm ion}$, number of produced ionising
photons per unit UV luminosity; e.g.][]{Matthee2017LyC}. $\chi_{\rm Ly\alpha}$ is typically computed with large spectroscopic samples, with $\chi_{\rm Ly\alpha}$ being the ratio between the number of galaxies with Ly$\alpha$ emission detected above some Ly$\alpha$ EW threshold and the total number of probed galaxies \citep[see e.g.][]{Stark2010}. $\chi_{\rm Ly\alpha}$ is found to be higher for galaxies fainter in the rest-frame UV \citep[M$_{\rm UV}$, e.g.][]{Pentericci2011}, implying such galaxies have higher escape fraction of Ly$\alpha$ photons and/or have a higher $\xi_{\rm ion}$ \citep[e.g][]{Maseda2020}. This can also be linked with faint M$_{\rm UV}$ galaxies having higher Ly$\alpha$ EW \citep[see e.g.][]{Shimizu2011,Kusakabe2018} and thus being more susceptible to being picked as LAEs, although such trend could also be a consequence of selection effects or survey limits \citep[see e.g.][]{Nilsson2009, Ando2006, Zheng2014,Hashimoto2017}. Some studies report no strong correlation between $\chi_{\rm Ly\alpha}$ and M$_{\rm UV}$ \citep[][]{Kusakabe2020} and attribute the typical high $\chi_{\rm Ly\alpha}$ of faint M$_{\rm UV}$ galaxies to selection biases in Lyman break galaxy (LBG) samples, which are biased towards selecting sources with high Ly$\alpha$ EW, as strong Ly$\alpha$ emission will boost the photometry and enhance the Lyman break, making such sources easier to detect.

Alternatively, $\chi_{\rm Ly\alpha}$ could in principle be inferred from the ratio between luminosity functions (LF, number density per luminosity bin vs luminosity) of Ly$\alpha$-selected and UV continuum-selected samples. The UV LF of continuum-selected galaxies has been extensively constrained in multiple studies up to $z\sim10$ \citep[e.g.][]{Steidel1999,Arnouts2005,Sawicki2006,Bouwens2015,Finkelstein2015,Alavi2016,Mehta2017,Ono2018}. The characteristic number density ($\Phi^*$) is found to decrease with an increasing redshift, from $\log_{10}(\Phi^*/$Mpc$^{-3})\sim-2.5$ at $z\sim2$ \citep{Reddy2009} to $\sim-3.5$ at $z\sim6$ \citep{Bouwens2015}. The UV LF of LAEs has also been probed by multiple studies \citep[see e.g.][]{Hu2004,Shimasaku2006,Ouchi2008}, targeting volumes of up to $\sim10^6$\,Mpc$^3$. \cite{Ouchi2008} found no evolution of the UV LF of LAEs at $z\sim3-4$, but an increase of UV bright LAEs at $z=5.7$. It is important to establish whether such evolutionary trends hold for much larger volumes ($\sim10^8$\,Mpc$^3$) and larger samples of LAEs.

Furthermore, it is important to establish how LAEs contribute to the total mass budget of galaxies. LAEs are typically low stellar mass galaxies, but can span a wide range of stellar masses, with some LAEs being very massive \citep[$\rm>10^{10}\,M_\odot$, e.g.][]{Finkelstein2009}. The stellar mass function (SMF) of continuum-selected galaxies has been well studied up to $z\sim4$ \citep[see e.g.][]{Pozzetti2010,Mortlock2011,Santini2012,Ilbert2013,Muzzin2013}. The SMF of continuum-selected galaxies is found to shift to lower number densities ($\Phi^*$ decrease) with increasing redshift, from $\log_{10}(\Phi^*/$Mpc$^{-3})\sim-3.3$ at $z=2-2.5$ to $\sim-5.0$ at $z=3-4$ \citep{Muzzin2013}. For LAEs, our understanding of the SMF is very limited as most studies are only able to determine stellar masses of stacks of the population \citep[e.g.][]{Kusakabe2018}. Estimating stellar masses of high-redshift LAEs is challenging due to near-infrared (NIR) coverage typically not being deep enough. Recent programs such as UltraVISTA \citep{McCracken2012} DR4 provide ultra-deep NIR imaging which can be used to better constrain the spectral energy distribution of high-redshift galaxies. Measurements of the stellar mass of individual galaxies in large samples spanning wide redshift ranges can significantly improve our view on the evolution of LAEs and how they compare with more typical galaxy samples.

In this work, we use a uniformly selected sample of $\sim4000$ LAEs \citep[SC4K,][]{Sobral2018} to compute UV LFs and SMFs in the wide redshift range $z\sim2-6$. We use the publicly available catalogues from \cite{Calhau2020} which identify AGN candidates in the SC4K sample using X-ray and radio measurements and the publicly available catalogues from \cite{Santos2020} which has measured the UV luminosity and stellar mass of LAEs in the SC4K sample. By comparing the luminosity and stellar mass density of LAEs with measurements of continuum-selected galaxies, we can infer how representative LAEs are of the overall population of galaxies at different redshifts.

This paper is structured as follows: in Section \ref{sec:sample}, we present the SC4K sample of LAEs, together with some galaxy properties derived in previous works. We present our methodology to derive UV LFs and SMFs in Section \ref{sec:methods}. We present and discuss our results in Section \ref{sec:results}, probing the evolution of the UV LF and SMF parameters across time, as well as estimating the evolution of $\Phi_{\rm LAE}/\Phi_{\rm LBG}$ (proxy of $\chi_{\rm Ly\alpha}$) and the luminosity and stellar mass densities. We present our conclusions in Section \ref{sec:conclusions}. Throughout this work, we use a $\Lambda$CDM cosmology with H$_0 = 70$\,km\,s$^{-1}$\,Mpc$^{-1}$, $\Omega_{\rm M} = 0.3$ and $\Omega _\Lambda = 0.7$. All magnitudes in this paper are presented in the AB system \citep{Oke1983} and we use a \citet{Chabrier2003} initial mass function (IMF).

\section{Sample and properties} \label{sec:sample}

\subsection{SC4K sample of LAEs} \label{subsec:sample}

%
%
\begin{table}
\setlength{\tabcolsep}{2.8pt}
\begin{center}
\caption{Overview of the SC4K sample of LAEs used in this study (summary table of \citealt{Sobral2018} and \citealt{Santos2020}). Given values are the median of all measurements for each galaxy property, with the errors being the 16th and 84th percentile of the distribution. (1) LAE selection filter \citep{Sobral2018}; (2) Redshift range the filter is sensitive to Ly$\alpha$ emission, based on the filter FWHM; (3) Number of LAEs; (4) Co-moving volume probed by each filter; (5) Median-likelihood stellar mass parameter from SED fitting, see \S\ref{subsec:mstar}; (6) UV magnitude computed by integrating the SED at $\lambda_0=1500$\,{\AA}, see \S\ref{subsec:muv}.} \label{tab:overview}
\begin{tabular}{c | ccccc}
\hline
(1)& (2) & (3) & (4) & (5) & (6)\\
\multicolumn{1}{c|}{Filter} &
\multicolumn{1}{c|}{Ly$\alpha\,z$} &
\multicolumn{1}{c|}{\# LAEs} &
\multicolumn{1}{c|}{Volume} &
\multicolumn{1}{c|}{M$_\star$} &
\multicolumn{1}{c|}{M$_{\rm UV}$} \\
& & & ($\times10^6$  & (log$_{10}\,$ & (AB)\\
& & & Mpc$^3$)  & (M$_\star$/M$_{\odot}$)) & \\
\hline
NB392 & $2.20-2.24$ & 159& 0.6& $9.5^{+0.5}_{-0.6}$ & $-19.6^{+1.0}_{-0.6}$\\
IA427 & $2.42-2.59$ & 741& 4.0& $9.2^{+0.5}_{-0.5}$ & $-19.7^{+0.6}_{-0.6}$ \\
IA464 & $2.72-2.90$ & 311& 4.2&$9.1^{+0.6}_{-0.3}$ & $-20.2^{+0.5}_{-0.5}$ \\
IA484 & $2.89-3.08$ & 711& 4.3& $9.0^{+0.7}_{-0.3}$ & $-20.0^{+0.6}_{-0.7}$ \\
NB501 & $3.08-3.16$ & 45& 0.9& $9.6^{+0.4}_{-0.5}$ & $-20.4^{+1.1}_{-0.8}$ \\
IA505 & $3.07-3.26$ & 483 & 4.3 & $9.4^{+0.5}_{-0.5}$ & $-20.2^{+0.6}_{-0.6}$ \\
IA527 & $3.23-3.43$ & 641 & 4.5 & $9.4^{+0.6}_{-0.6}$ & $-20.2^{+0.5}_{-0.6}$ \\
IA574 & $3.63-3.85$ & 98 & 4.9 & $9.3^{+0.7}_{-0.2}$ & $-20.8^{+0.5}_{-0.4}$ \\
IA624 & $4.00-4.25$ & 142 & 5.2 & $9.2^{+0.5}_{-0.5}$ & $-20.5^{+0.5}_{-0.6}$ \\
IA679 & $4.44-4.72$ & 79 & 5.5 & $9.5^{+0.8}_{-0.3}$ & $-21.2^{+0.6}_{-0.5}$ \\
IA709 & $4.69-4.95$ & 81 & 5.1 & $9.4^{+0.5}_{-0.3}$ & $-21.1^{+0.5}_{-0.4}$ \\
NB711 & $4.83-4.89$ & 78 & 1.2 & $9.7^{+0.6}_{-0.6}$ & $-20.9^{+0.5}_{-0.8}$ \\
IA738 & $4.92-5.19$ & 79 & 5.1 & $9.6^{+0.7}_{-0.3}$ & $-21.3^{+0.4}_{-0.7}$ \\
IA767 & $5.17-5.47$ & 33 & 5.5 & $9.7^{+0.3}_{-0.4}$ & $-21.6^{+0.4}_{-0.5}$ \\
NB816 & $5.65-5.75$ & 192 & 1.8 & $9.9^{+0.4}_{-0.5}$ & $-21.4^{+0.6}_{-0.6}$ \\
IA827 & $5.64-5.92$ & 35 & 4.9 & $9.9^{+0.6}_{-0.4}$ & $-22.0^{+0.8}_{-1.0}$ \\
\hline
Full SC4K & $2.20-5.92$ & 3908 & 62.0 & $9.3^{+0.6}_{-0.5}$ & $-20.2^{+0.7}_{-0.8}$\\
\hline
\end{tabular} \label{table:overview}
\end{center}
\end{table}

%
%
\begin{figure*}
\begin{tabular}{ll}
  \centering
  \includegraphics[width=0.487\textwidth]{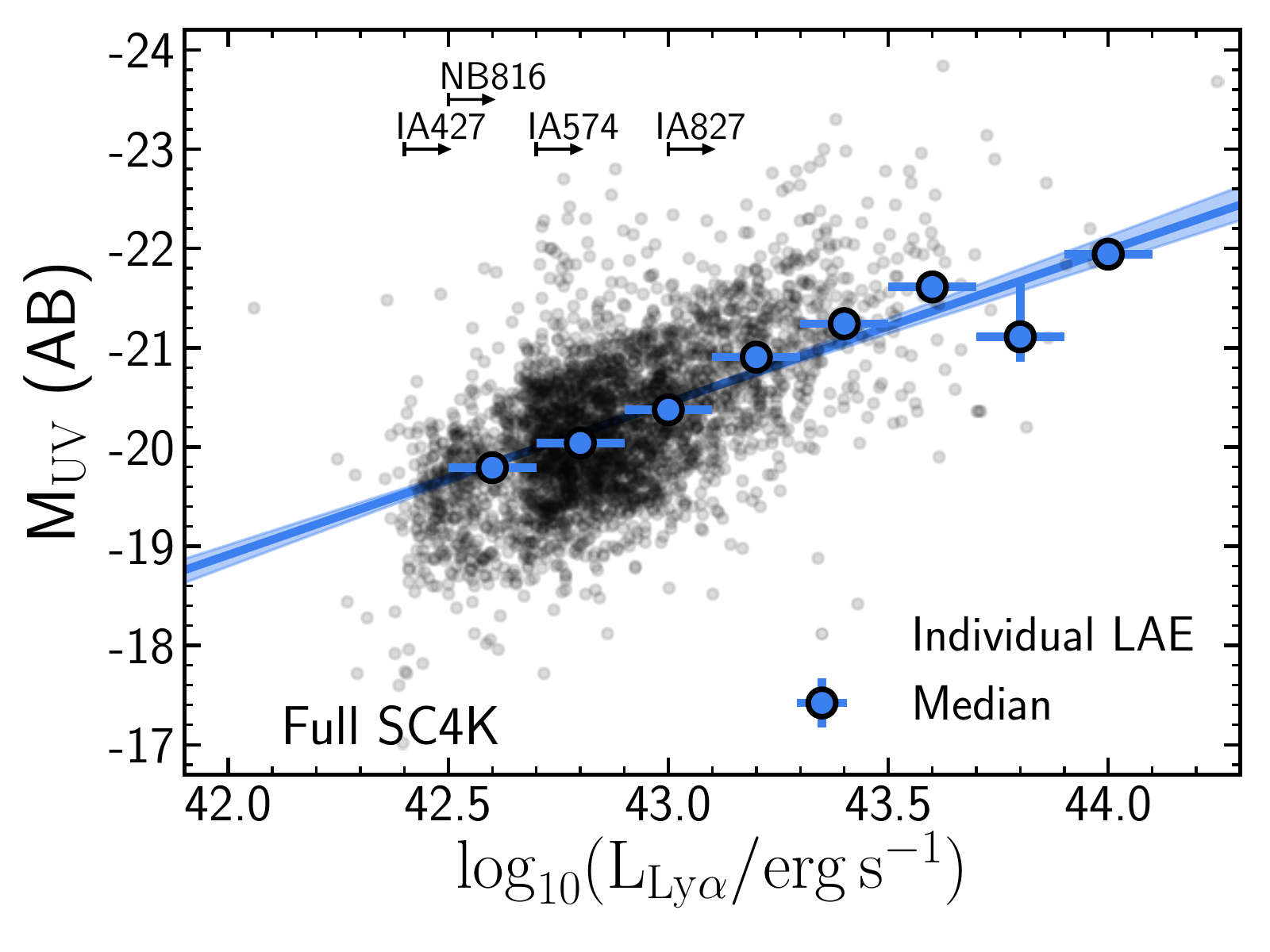}
  &
  \includegraphics[width=0.487\textwidth]{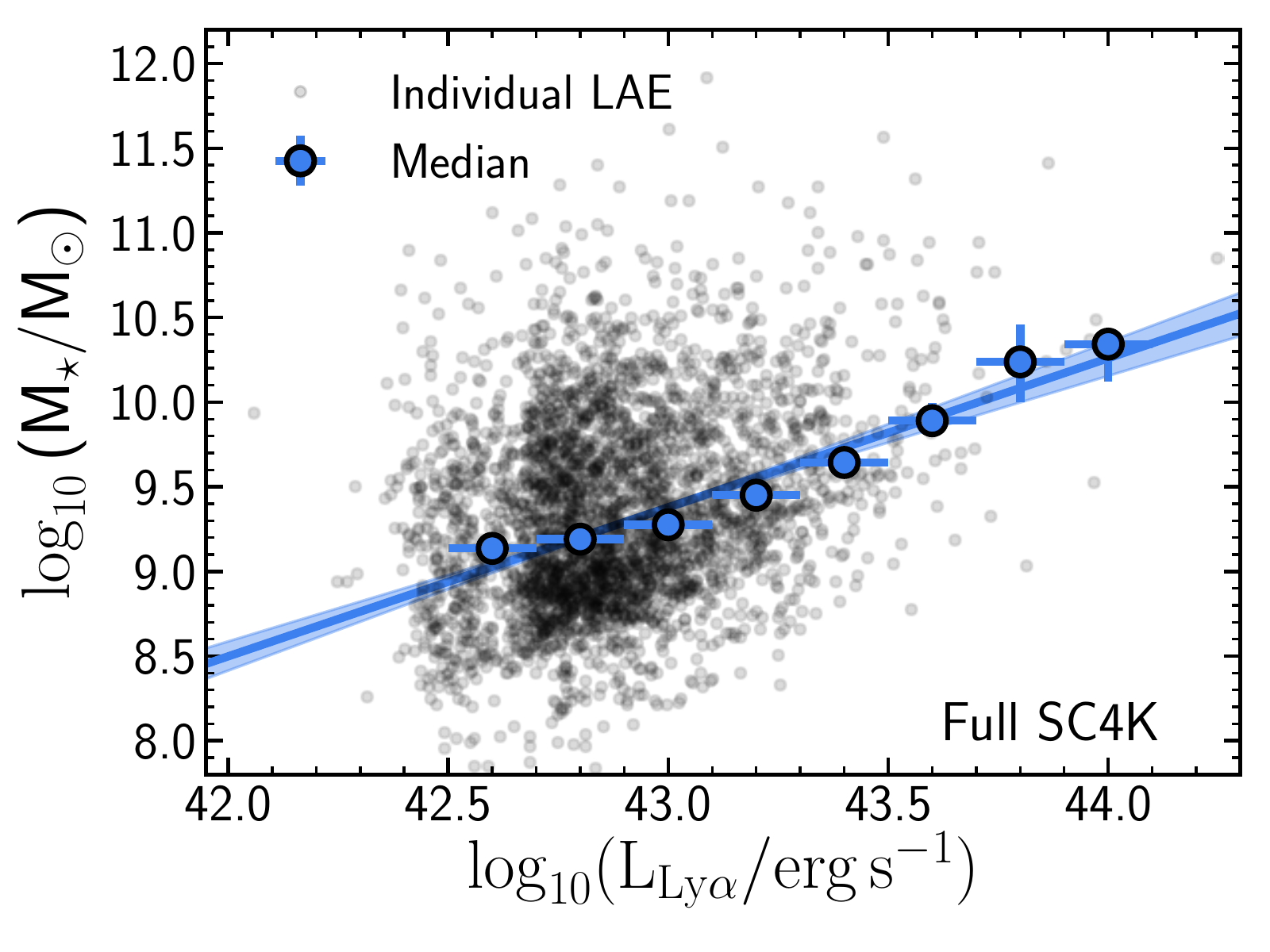}  
\end{tabular}
  \caption{{\it Left:} M$_{\rm UV}$ dependence on L$_{\rm Ly\alpha}$ within our sample of LAEs. Individual measurements are plotted as scatter in the background. We calculate the median M$_{\rm UV}$ per $\rm L_{Ly\alpha}$ bin (blue circles), with the error bars being the 16th and 84th percentile of the M$_{\rm UV}$ distribution divided by $\sqrt{\rm N}$, with N being the number of sources inside the bin. Bins are defined with 0.2 bin width, starting at $\log_{10} (\rm L_{Ly\alpha}/erg\,s^{-1})=42.5$, which corresponds to the 3$\sigma$ L$_{\rm Ly\alpha}$ limit for the MB at $z=2.5$. The blue shaded contour is the 16th and 84th percentiles of 1000 iterations of fits, obtained by perturbing the median bins within their asymmetric error bars. We find M$_{\rm UV}$ and L$_{\rm Ly\alpha}$ to be well correlated (best-fit M$_{\rm UV}=-1.6_{-0.3}^{+0.2}\log_{10} (\rm L_{Ly\alpha}/erg\,s^{-1})+47_{-11}^{+12}$) in our sample of LAEs, with bright M$_{\rm UV}$ typically corresponding to bright L$_{\rm Ly\alpha}$, but with an important scatter. There is a clear and gradual median brightening at $\log_{10} (\rm L_{Ly\alpha}/erg\,s^{-1})=42.5-43.5$, from -19.8 to -21.4. The higher number of sources above $\log_{10} (\rm L_{Ly\alpha}/erg\,s^{-1}) \approx 42.7$ (also observed in the right panel) is a consequence of flux limit differences between narrow and medium bands. For reference, we show the 3$\sigma$ L$_{\rm Ly\alpha}$ limits for the IA427 (MB, $z=2.5$), IA574 (MB, $z=3.7$), IA827 (MB, $z=5.7$) and NB816 (NB, $z=5.7$) samples. {\it Right:} Same but for stellar mass (M$_\star$). We also find a correlation between M$_\star$ and L$_{\rm Ly\alpha}$ (best-fit $\log_{10} (\rm M_\star/M_\odot)=0.9_{-0.1}^{+0.1}\log_{10} (\rm L_{Ly\alpha}/erg\,s^{-1})-28_{-3.8}^{+4.0}$), which is shallower than the correlation found for M$_{\rm UV}$, revealing how a stellar mass selection and a Ly$\alpha$ selection can differ. The median evolution is less evident than in the left panel, with the median $\log_{10} (\rm M_\star/M_\odot)$ only increasing by 0.2 in the luminosity range $\log_{10} (\rm L_{Ly\alpha}/erg\,s^{-1})=42.5-43.0$.}
  \label{fig:lya_vs_others}
\end{figure*}

The public SC4K sample of LAEs \citep[Slicing COSMOS with 4k LAEs,][]{Sobral2018} consists of 3908 LAEs selected with 12+4 medium+narrow band (MB+NB) filters (see Table \ref{tab:overview} for an overview) over the 2 deg$^2$ of the COSMOS field \citep{Capak2007,Scoville2007,Taniguchi2015}. For full details on the selection criteria applied, we refer the reader to \cite{Sobral2018}. Briefly, LAEs are selected based on 1) Ly$\alpha$ EW$_0>50$\,\AA\,(25\,\AA\,for NBs and 5\,\AA\, for NB392); 2) significant excess emission \citep[$\Sigma>3$; see][]{Bunker1995}; 3) colour break blueward of the Ly$\alpha$ emission; 4) exclusion of sources with strong red colours (prevents lower redshift interlopers with strong Balmer breaks); 5) full visual inspection to remove spurious detections.

Multiple studies have used the SC4K sample to derive properties of LAEs. \cite{PaulinoAfonso2018} and \cite{Shibuya2019} find small UV sizes with little evolution from $z\sim2$ to $z\sim6$. Clustering analysis reveals dark matter halo masses strongly depend on the Ly$\alpha$ luminosity \citep[L$_{\rm Ly\alpha}$,][]{Khostovan2019}. \cite{Calhau2020} analysed X-ray and radio data on the COSMOS field and measured a low  ($<10\%$) overall AGN fraction, dependent on L$_{\rm Ly\alpha}$, significantly increasing with increasing luminosity and approaching $100\%$ at L$_{\rm Ly\alpha}>10^{44}$ erg s$^{-1}$. SED fitting from \cite{Santos2020} shows that SC4K LAEs are typically very blue ($\beta=-2.1$), low mass (M$_\star=10^{9.3}$\,M$_\odot$), and above the star-forming Main Sequence at $z<4$ and M$_\star<10^{9.5}$\,M$_\odot$. SC4K sources are also the prime focus of follow-up spectroscopic observations focusing on studying primeval galaxies \citep{Amorin2017}.

\subsubsection{X-ray and radio AGN in SC4K} \label{sec:sample_AGNs}

The SC4K sample includes 254 LAEs detected in X-ray and 120 detected in radio (56 in both), resulting in 318 AGN candidates \citep{Calhau2020} out of 3705 SC4K LAEs with X-ray or radio coverage. Following the same methodology as \cite{Santos2020}, we classify these sources as AGNs since pure star-forming processes would require extremely high SFRs ($\gtrsim1000$\,M$_\odot$\,yr$^{-1}$) to be detectable at such wavelengths and redshifts. Throughout this work, SC4K AGNs are removed from any fitting/binning and median values in tables, except in Figures \ref{fig:grid_muv_lf} and \ref{fig:lf_full} (left panel), where we show the number densities of AGN LAEs.

\subsubsection{Redshift binning} \label{subsec:binning}

In addition to analysing the properties of LAEs from specific selection filters, we group filters with similar central wavelengths to analyse specific redshift bins in a more statistically robust way. We use a grouping scheme similar to \cite{Sobral2018} and \cite{Santos2020}:
\begin{itemize}
\item $z=2.5\pm0.1$ (IA427);
\item $z=3.1\pm0.4$ (IA464, IA484, NB501, IA505, IA527);
\item $z=3.9\pm0.3$ (IA574, IA624);
\item $z=4.7\pm0.2$ (IA679, IA709, NB711);
\item $z=5.4\pm0.5$ (IA738, IA767, NB816, IA827).
\end{itemize}

In this work, we include NBs in the redshift bins (even though they typically reach fainter Ly$\alpha$ luminosities) as we perform Ly$\alpha$ luminosity cuts to ensure the samples are directly comparable (see \S\ref{sec:muv_vary} and \S\ref{sec:smf_vary}).

%
%
\begin{figure*}
\begin{tabular}{ll}
  \centering
  \includegraphics[width=0.487\textwidth]{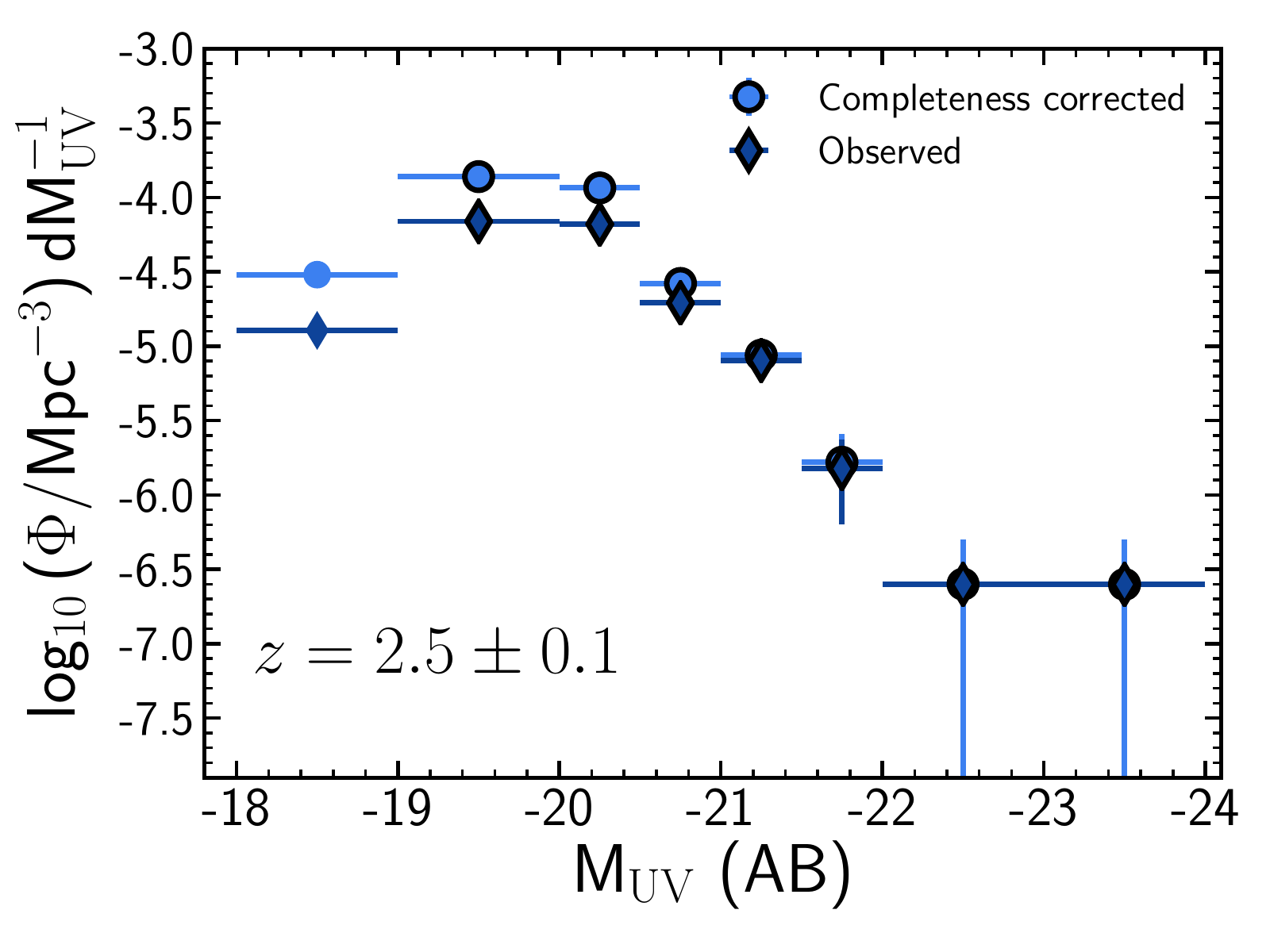}
  &
  \includegraphics[width=0.487\textwidth]{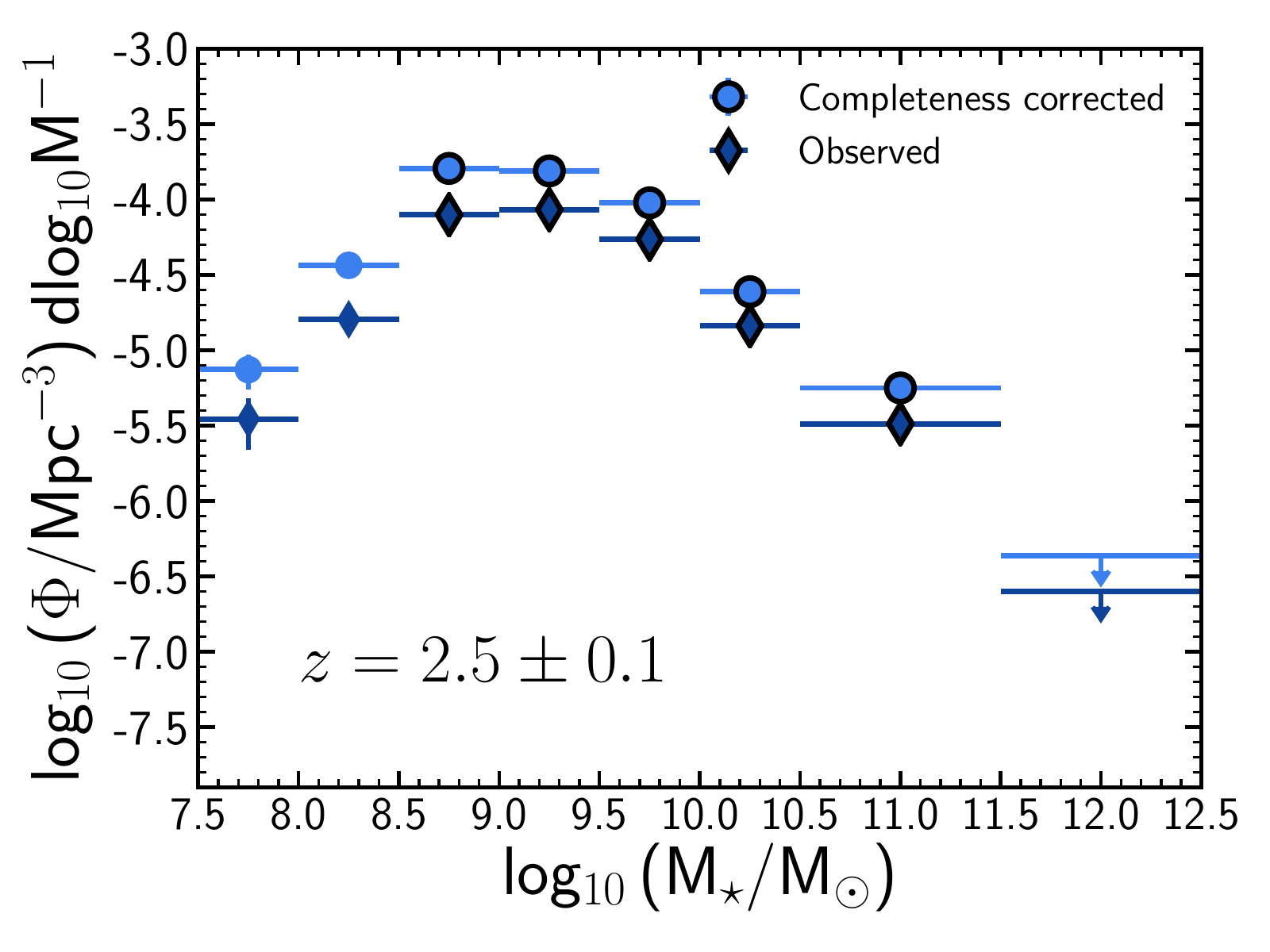}  
\end{tabular}
  \caption{{\it Left:} The rest-frame UV LF for the $z=2.5$ (IA427) sample of LAEs. We show the luminosity values before (dark blue diamonds) and after (blue circles) applying the completeness correction. The completeness correction is based on Ly$\alpha$ flux (selection criteria). Since M$_{\rm UV}$ and L$_{\rm Ly\alpha}$ typically correlate (see Fig. \ref{fig:lya_vs_others}, left panel), the completeness corrections are larger for the faintest M$_{\rm UV}$ bins. {\it Right:} Same but for the SMF. As the correlation between L$_{\rm Ly\alpha}$ and M$_\star$ is shallower (see Fig. \ref{fig:lya_vs_others}, right panel), completeness corrections are not a strong function of stellar mass for a specific Ly$\alpha$ cut.}
  \label{fig:completeness}
\end{figure*}

\subsection{Spectral energy distribution and properties of SC4K LAEs}
Spectral energy distribution (SED) fitting of the full SC4K sample is presented in \cite{Santos2020}. Briefly, SED-fitting is done using the publicly available SED-fitting code {\sc magphys}\footnote{http://www.iap.fr/magphys/} \citep{daCunha2008,daCunha2012} with the high-redshift extension \cite[see][]{daCunha2015}, which models stellar and dust emission from galaxies. We obtain photometric measurements from publicly available imaging, taken with 34 rest-frame UV-FIR filters in the COSMOS field  \citep{Capak2007,McCracken2012,Steinhardt2014,Sanders2007,Lutz2011,Oliver2012}. As the SED-fitting code does not include nebular emission, we exclude the NB or MB with observed Ly$\alpha$ emission from the SED-fitting. Derived parameters are the median-likelihood parameters obtained by comparing modelled SEDs with libraries of galaxies at similar redshift. {\sc magphys} uses dust attenuation models from \cite{Charlot2000} and the stellar population synthesis model from \cite{Bruzual2003} with a \cite{Chabrier2003} IMF (range 0.1-100 M$_\odot$). The prescription of \cite{Madau1995} is used to model the intergalactic medium (IGM).

In this work, we will focus on two SED-derived properties: rest-frame UV luminosity (M$_{\rm UV}$) and stellar mass (M$_\star$). We use the public catalogues provided by \cite{Santos2020}, which contain coordinates, photometry and derived galaxy properties for the full SC4K sample of LAEs.

\subsubsection{Rest-frame UV luminosity {\rm (M$_{\rm UV}$)}} \label{subsec:muv}

The UV luminosity of a galaxy (M$_{\rm UV}$) can be used as a tracer of recent star-formation on $\sim100$\,Myr timescales \citep[e.g.][]{Boselli2001,Salim2009}. M$_{\rm UV}$ is computed in \cite{Santos2020} by integrating the best-fit SEDs at rest-frame $\lambda_0=1400-1600$\,\AA. The median of the SC4K sample is M$_{\rm UV}=-20.2^{+0.7}_{-0.8}$ (Table \ref{tab:overview}), which corresponds to $0.09\times$L$_{z=3}^*$ \citep{Steidel1999}.

Similarly but for shorter timescales, Ly$\alpha$ emission also traces recent star-formation, due to being a tracer of Lyman-Continuum \citep[e.g.][]{SobralMatthee2019} like H$\alpha$ \citep[][]{Kennicutt1998}. As the massive young stars responsible for producing the UV continuum also produce the ionising photons that lead to Ly$\alpha$ emission, we can expect these two properties to be related. For our sample of LAEs, we observe that these two properties are typically correlated (see Fig. \ref{fig:lya_vs_others}, left panel), with the median M$_{\rm UV}$ significantly brightening from -19.8 to -21.4 in the luminosity range $\log_{10} (\rm L_{Ly\alpha}/erg\,s^{-1})=42.5-43.5$. We compute a best-fit of M$_{\rm UV}=-1.6_{-0.3}^{+0.2}\log_{10} (\rm L_{Ly\alpha}/erg\,s^{-1})+47_{-11}^{+12}$ from the median distribution. However, Ly$\alpha$ luminosity (L$_{\rm Ly\alpha}$) does not necessarily translate M$_{\rm UV}$ and vice-versa (see the scatter around M$_{\rm UV}=-20$ and see discussion in \citealt{Matthee2017spectra} and \citealt{Sobral2018Spectra}). This is also made evident from LBG samples, where there are bright M$_{\rm UV}$ sources with no significant Ly$\alpha$ detection, as shown by the Ly$\alpha$ fraction \citep[e.g.][]{Stark2010,Pentericci2011,Arrabal2018,Kusakabe2020}.

\subsubsection{Stellar Mass {\rm (M$_\star$)}} \label{subsec:mstar}

The shape and normalisation of an SED is a reflection of the content of stars in a galaxy, thus its total mass of stars (stellar mass, M$_\star$) can be derived by fitting and modelling the SED. LAEs typically have low M$_\star$ but there is an important diversity within the population. The median of the SC4K sample of LAEs is computed in \cite{Santos2020} using {\sc magphys}: log$_{10}\,$(M$\rm_\star/M_{\odot})=9.3^{+0.6}_{-0.5}$ (Table \ref{tab:overview}), which corresponds to $0.006\times$\,M$_{\star,z=3-4}^*$ \citep{Muzzin2013}. We find the median M$_\star$ and L$_{\rm Ly\alpha}$ to be correlated (see Fig. \ref{fig:lya_vs_others}, right panel), with best-fit $\log_{10} (\rm M_\star/M_\odot)=0.9_{-0.1}^{+0.1}\log_{10} (\rm L_{Ly\alpha}/erg\,s^{-1})-28_{-3.8}^{+4.0}$, but with a significant scatter of individual detections. This relation is shallower than the one measured between M$_{\rm UV}$ and L$_{\rm Ly\alpha}$, with a more modest increase: the median $\log_{10} (\rm M_\star/M_\odot)$ only increases by 0.2 in the luminosity range $\log_{10} (\rm L_{Ly\alpha}/erg\,s^{-1})=42.5-43.0$. We note that the SED-fitting used to derive M$_\star$ does not include nebular emission, so the two properties are independently derived. Additionally, there is an anti-correlation between M$_\star$ and Ly$\alpha$ EW$_0$, and thus Ly$\alpha$ escape fraction of LAEs \citep{Santos2020}.

\section{Luminosity and stellar mass functions} \label{sec:methods}

In this section, we present our methodology and computations to derive UV LFs and SMFs for our sample of $\sim4000$ LAEs at well-defined redshift intervals between $z\sim2$ and $z\sim6$. 

%
%
\begin{figure*}
  \centering
  \includegraphics[width=\textwidth]{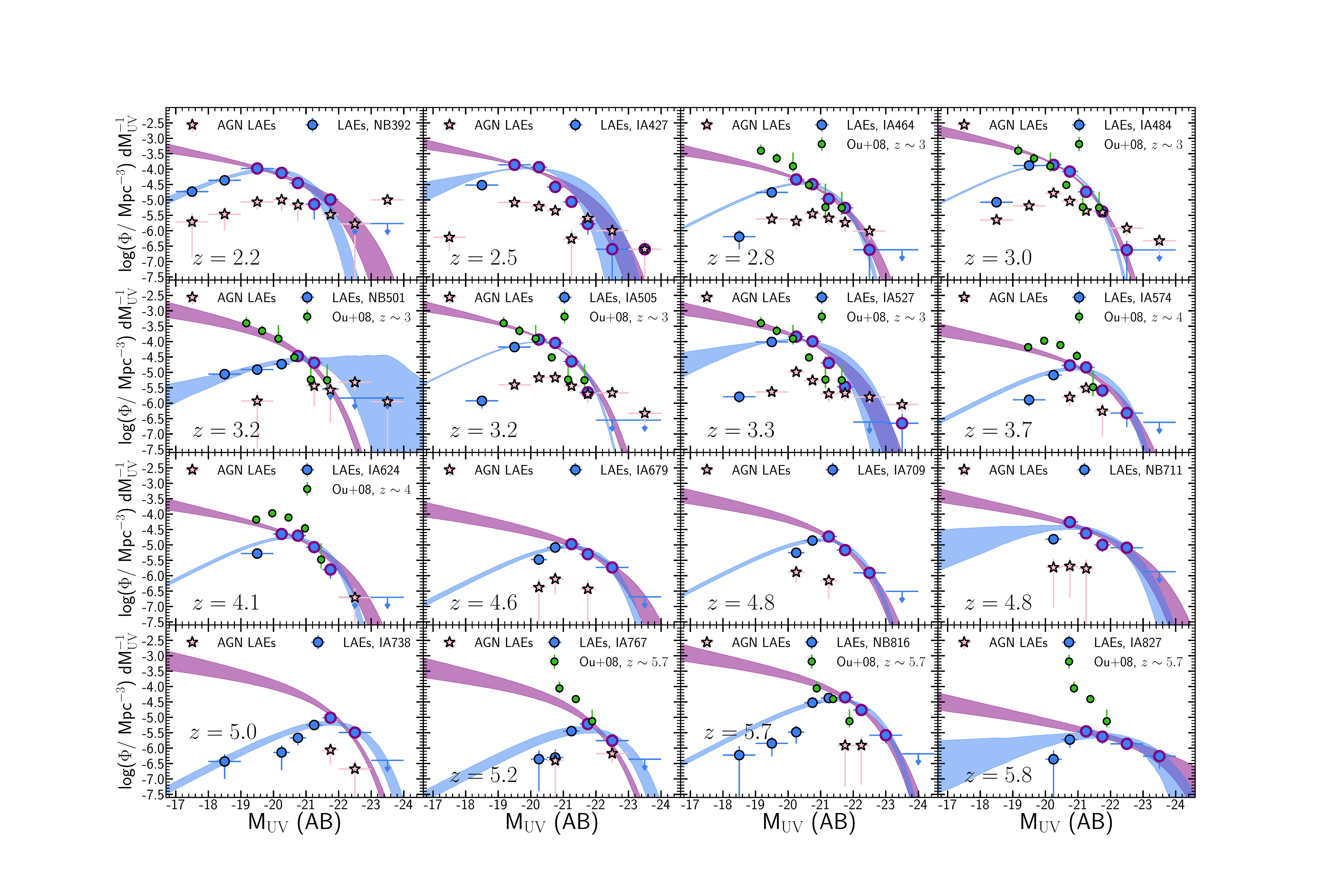}
  \caption{The rest-frame UV LF (blue circles) for each of the 16 individual selection filters in this study, without any Ly$\alpha$ flux cut and excluding AGNs (see \S\ref{sec:sample_AGNs}). Luminosity bins brighter than the peak of number densities are marked with a purple edge colour. The blue contours are the 16th and 84th percentile of multiple iterations of fits to the luminosity bins, obtained by perturbing the luminosity bins within their error bars (see \S\ref{subsec:perturb_fits}) for the full UV luminosity range. The purple contours represent the same but only fitting the points above the number density peak. We compare our results with those of \citet{Ouchi2008} at $z\sim3, 4, 5.7$, finding a good agreement, with the offset at $z=5.8$ being easily explained by differences in Ly$\alpha$ flux limits. We show the number densities of AGNs (pink stars), which predominantly dominate the bright M$_{\rm UV}$  regime ($<-22$) at $z<4$, often having higher number densities than non-AGN LAEs. At $z>4$ there is significantly less AGN LAEs, composing only a small fraction of LAEs at all M$_{\rm UV}$ ranges.}
  \label{fig:grid_muv_lf}
\end{figure*}

%
%
\begin{figure*}
  \centering
  \includegraphics[width=\textwidth]{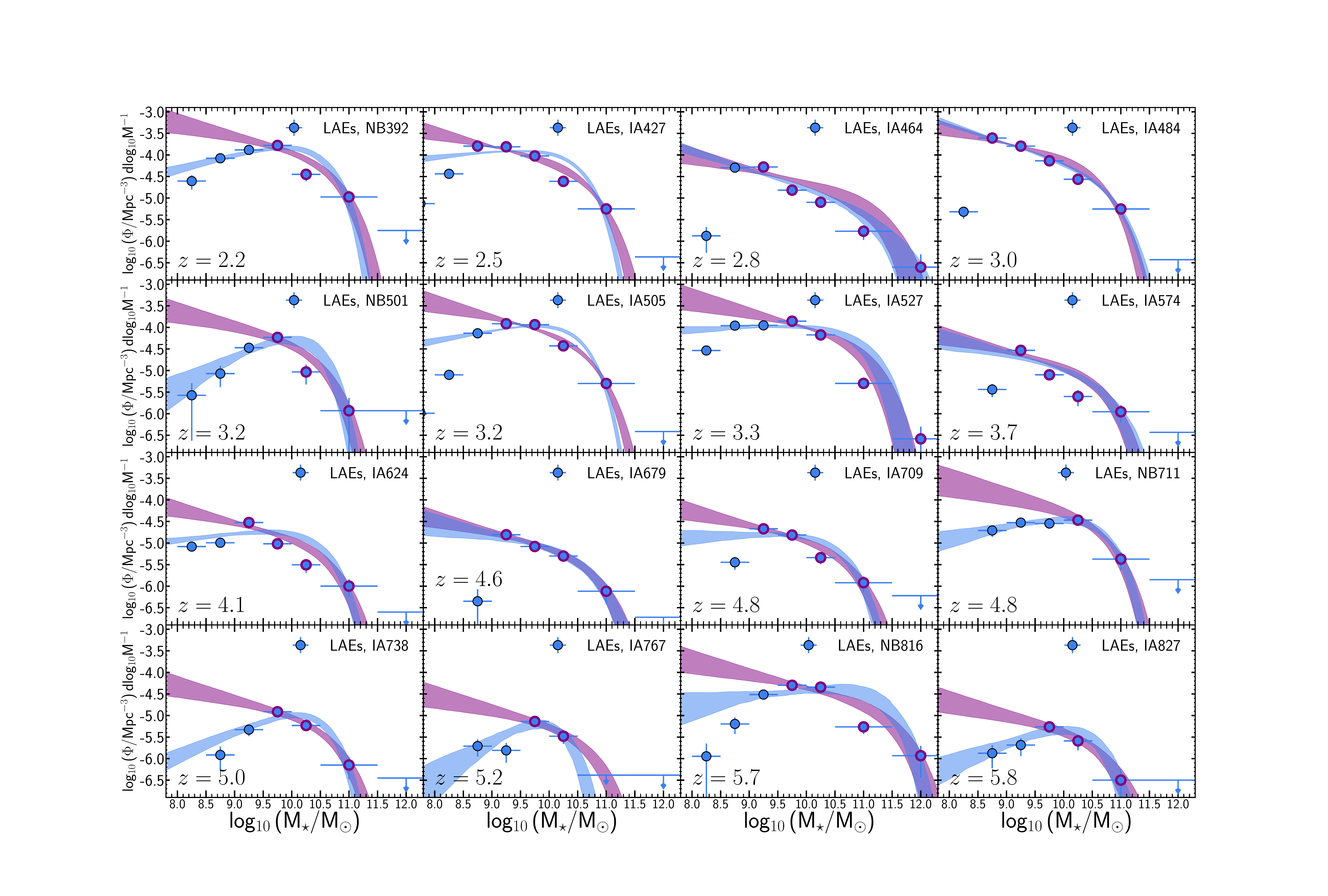}
  \caption{The SMF for each of the 16 individual selection filters. Stellar mass bins are shown as blue circles. Stellar mass bins more massive than the peak of number densities are marked with a purple edge colour. The blue contours are the 16th and 84th percentile of multiple iterations of fits to the stellar mass bins, obtained by perturbing the stellar mass bins within their error bars (see \S\ref{subsec:perturb_fits}) for the full stellar mass range. The purple contours represent the same but only fitting the points above the number density peak. Candidate AGN are removed from the analysis here (see \S\ref{sec:sample_AGNs})}
  \label{fig:grid_smf}
\end{figure*}

\subsection{Determining the luminosity/mass functions}

We measure the number densities of well-defined M$_{\rm UV}$ and M$_\star$ bins which we use to construct the UV LF and SMF. We choose bin widths depending on M$_{\rm UV}$ and M$_\star$, as the most (and least) luminous and massive bins have fewer sources. We define bins with width 0.5 dex in the range $-22<{\rm M}_{\rm UV}<-20$ (M$_{\rm UV}>-22.5$ for the deeper NB816) and $7<\log_{10} ({\rm M_\star/M_\odot})<10.5$ and 1 dex outside these ranges, where the number densities are the lowest.  We use Poissonian errors for any individual LF realisation.

The number density of a luminosity bin is given by:

\begin{eqnarray}
\log_{10}(\phi_j)=\log_{10}\left(\frac{1}{\mathrm{d}\log_{10}\mathrm{L}}\frac{N_j}{V}\right),
\label{eq:bins}
\end{eqnarray}

\noindent where $\phi_j$ is the number density of a bin $j$, $N_j$ is the number of sources within d$\log_{10}$L of $j$, and $V$ is the volume probed by the NBs or MBs for that specific bin (see Table \ref{table:overview}), which is computed from the redshift range that each filter is sensitive to Ly$\alpha$ emission.
 
\subsection{Completeness correction} \label{subsec:completeness}

Faint sources and those with low Ly$\alpha$ EW may be missed by our selection criteria, leading to an underestimation of number densities. We estimate completeness corrections based on Ly$\alpha$ line flux (same corrections used for the Ly$\alpha$ LFs in \citealt{Sobral2018}; full details therein) and apply them to the UV LFs and the SMFs of our sample of LAEs. Briefly, for each NB or MB, we obtain a sample of high-redshift non-line-emitters by applying the same colour break we used to target the Lyman break in our LAEs and by selecting sources with photometric redshifts \citep[obtained from][]{Laigle2016} $\pm0.2$ the redshift range given in Table \ref{tab:overview}. The distribution of MB/NB magnitudes in this non-line-emitter sample is similar to the distribution of LAEs, with a tail of $2-5\%$ brighter sources. The non-line-emitter sample is thus slightly brighter than the sample of LAEs and provides a sightly conservative estimation of the completeness corrections. For each non-line-emitter sample, flux is incrementally added to the NB or MB and BB \cite[see][Table 3 therein]{Sobral2018}. By reapplying our selection criteria after each step, we determine the fraction of galaxies which are picked as emitters per Ly$\alpha$ luminosity value.  We only consider sources with $>30\%$ completeness.

We apply completeness corrections to each LAE individually, based on their observed Ly$\alpha$ flux, and not their M$_{\rm UV}$ or M$_\star$. In Fig. \ref{fig:completeness}, we show M$_{\rm UV}$ and M$_\star$ number densities for $z=2.5$ (IA427) LAEs, before and after completeness corrections. We note that the completeness correction is based on Ly$\alpha$ flux and thus larger for fainter LAEs but not necessarily correlated with other properties. Since M$_{\rm UV}$ and L$_{\rm Ly\alpha}$ typically correlate (see Fig. \ref{fig:lya_vs_others}, left panel), the corrections will typically be smaller for LAEs which are brighter in M$_{\rm UV}$ (see Fig. \ref{fig:completeness}, left panel). Since the correlation between M$_\star$ and L$_{\rm Ly\alpha}$ is weak (see Fig. \ref{fig:lya_vs_others}, right panel), the corrections will be similar for the entire mass range (see Fig. \ref{fig:completeness}, right panel).

Including the completeness correction, applied to each source, Equation \ref{eq:bins} becomes:

\begin{eqnarray}
\log_{10}(\phi_j)=\log\left(\frac{1}{d\log_{10} L}\sum_i^{N_j}\frac{c_i}{V}\right),
\label{eq:bins_cor}
\end{eqnarray} 

\noindent where $c_i$ is the completeness correction for a source $i$.

For the luminosity or stellar mass bins with zero counts, we compute the upper limit as one source at the volume probed by the NB or MB, with the completeness correction equal to the total completeness correction applied to the previous luminosity or stellar mass bin.

\begin{table*}
\setlength{\tabcolsep}{5pt}
\caption{Best-fit Schechter parameters for the UV LF of LAEs from $z=2$ to $z=6$, for each of the individual selection filters and for different redshift bins (see \S\ref{subsec:binning}). The number of sources provided here is the number of sources included in the luminosity functions, i.e. non-AGN LAEs with available SEDs and with completeness corrections $>30\%$. We provide best fits for the two cases considered in this study: fit to the full UV luminosity range (blue in Fig. \ref{fig:grid_muv_lf}) and fit to the bins brighter than the number density peak (purple in Fig. \ref{fig:grid_muv_lf}). We provide the best set of parameters ($\alpha$, M$_{\rm UV}^*$ and $\Phi^*$) which minimise $\chi^2_{\rm red}$, with $\alpha$ being fixed for the latter case as it cannot be directly constrained. When $\chi^2_{\rm red}$ is very large, the errors should be interpreted with caution as the best parameters found still do not provide a good model. Additionally, M$_{\rm UV}^*$ is also fixed for the individual filters with less than three luminosity bins (although we perturb these parameters when exploring the uncertainties of the bins/fits, see \S\ref{subsec:perturb_fits}). For the redshift bins we also show the Schechter parameters when applying a $\log_{10} (\rm L_{Ly\alpha}/erg\,s^{-1}) \geq 43.0$ cut.} \label{tab:schechter_params_uv_lf}
\begin{tabular}{ccc | cccc | ccc}
\hline
  & & & \multicolumn{4}{c}{Full UV range}|  & \multicolumn{3}{c}{UV brighter than the peak}\\ 
   & & & \multicolumn{4}{c}{}  & \multicolumn{3}{c}{($\alpha_{\rm fix}=-1.5$)}\\
\hline
Redshift & \# Filters & \# Sources & $\log_{10}\Phi^*$ & M$_{\rm UV}^*$ & $\alpha$ & $\chi^2_{\rm red}$ & $\log_{10}\Phi^*$ & M$_{\rm UV}^*$ & $\chi^2_{\rm red}$\\
&  &  & (Mpc$^{-3}$) & (AB) & & &(Mpc$^{-3}$)&(AB)& \\
\hline
$2.2\pm0.1$ & 1 & 129 & $-3.57^{+0.03}_{-0.04}$ & $-19.78^{+0.16}_{-0.17}$ & $-0.7^{+0.1}_{-0.1}$ & 2.5 & $-4.16^{+0.11}_{-0.15}$ & $-21.15^{+0.26}_{-0.51}$ & 0.9 \\
$2.5\pm0.1$ & 1 & 519 & $-3.48^{+0.01}_{-0.01}$ & $-19.46^{+0.03}_{-0.04}$ & $-0.9^{+0.1}_{-0.1}$ & 33.9 & $-3.86^{+0.05}_{-0.06}$ & $-20.65^{+0.11}_{-0.15}$ & 14.0 \\
$2.8\pm0.1$ & 1 & 139 & $-4.04^{+0.03}_{-0.02}$ & $-20.02^{+0.06}_{-0.06}$ & $-1.5^{+0.1}_{-0.1}$ & 9.4 & $-4.19^{+0.09}_{-0.09}$ & $-20.95^{+0.15}_{-0.18}$ & 1.0 \\
$3.0\pm0.1$ & 1 & 565 & $-3.43^{+0.01}_{-0.02}$ & $-19.61^{+0.03}_{-0.03}$ & $-1.4^{+0.1}_{-0.1}$ & 36.8 & $-3.43^{+0.06}_{-0.06}$ & $-20.37^{+0.07}_{-0.09}$ & 3.6 \\
$3.2\pm0.1$ & 1 & 31 & $-4.11^{+0.61}_{-0.09}$ & $-21.09^{+0.49}_{-3.20}$ & $-0.3^{+0.1}_{-0.1}$ & 1.2 & $-3.65^{+0.04}_{-0.04}$ & $-20.37\,{\rm (fix)}$ & 1.7 \\
$3.2\pm0.1$ & 1 & 413 & $-3.59^{+0.01}_{-0.01}$ & $-19.82^{+0.03}_{-0.03}$ & $-0.7^{+0.1}_{-0.1}$ & 29.2 & $-3.58^{+0.05}_{-0.05}$ & $-20.58^{+0.08}_{-0.10}$ & 13.6 \\
$3.3\pm0.1$ & 1 & 565 & $-3.47^{+0.01}_{-0.01}$ & $-19.77^{+0.03}_{-0.03}$ & $-0.9^{+0.1}_{-0.1}$ & 36.4 & $-3.49^{+0.06}_{-0.05}$ & $-20.54^{+0.08}_{-0.09}$ & 12.3 \\
$3.7\pm0.1$ & 1 & 53 & $-4.45^{+0.03}_{-0.03}$ & $-20.51^{+0.09}_{-0.12}$ & $-1.3^{+0.1}_{-0.1}$ & 8.5 & $-4.45^{+0.11}_{-0.12}$ & $-21.17^{+0.16}_{-0.23}$ & 3.2 \\
$4.1\pm0.1$ & 1 & 116 & $-4.33^{+0.03}_{-0.02}$ & $-20.10^{+0.07}_{-0.08}$ & $-0.8^{+0.1}_{-0.1}$ & 13.4 & $-4.50^{+0.10}_{-0.12}$ & $-21.07^{+0.18}_{-0.27}$ & 2.9 \\
$4.6\pm0.1$ & 1 & 69 & $-4.64^{+0.03}_{-0.03}$ & $-20.84^{+0.06}_{-0.15}$ & $-1.3^{+0.1}_{-0.1}$ & 2.6 & $-4.85^{+0.09}_{-0.11}$ & $-21.90^{+0.14}_{-0.21}$ & 0.4 \\
$4.8\pm0.1$ & 1 & 50 & $-4.42^{+0.03}_{-0.03}$ & $-20.63^{+0.06}_{-0.08}$ & $-0.9^{+0.1}_{-0.1}$ & 6.3 & $-4.26^{+0.08}_{-0.08}$ & $-21.22^{+0.09}_{-0.12}$ & 0.5 \\
$4.8\pm0.1$ & 1 & 41 & $-4.04^{+0.07}_{-0.35}$ & $-21.30^{+0.31}_{-0.72}$ & $-0.7^{+0.1}_{-0.1}$ & 8.4 & $-4.44^{+0.10}_{-0.12}$ & $-22.10^{+0.20}_{-0.33}$ & 2.2 \\
$5.0\pm0.1$ & 1 & 29 & $-4.74^{+0.04}_{-0.04}$ & $-21.33^{+0.12}_{-0.20}$ & $-0.4^{+0.1}_{-0.1}$ & 3.0 & $-4.02^{+0.03}_{-0.03}$ & $-21.22\,{\rm (fix)}$ & 7.0 \\
$5.2\pm0.1$ & 1 & 17 & $-4.91^{+0.06}_{-0.05}$ & $-21.19^{+0.13}_{-0.24}$ & $0.0^{+0.1}_{-0.1}$ & 2.9 & $-4.25^{+0.04}_{-0.05}$ & $-21.22\,{\rm (fix)}$ & 3.0 \\
$5.7\pm0.1$ & 1 & 107 & $-3.98^{+0.03}_{-0.04}$ & $-21.09^{+0.09}_{-0.12}$ & $-1.0^{+0.1}_{-0.1}$ & 3.9 & $-3.84^{+0.08}_{-0.10}$ & $-21.67^{+0.10}_{-0.13}$ & 0.1 \\
$5.8\pm0.1$ & 1 & 14 & $-5.27^{+0.14}_{-0.82}$ & $-22.43^{+0.46}_{-2.53}$ & $-0.5^{+0.1}_{-0.1}$ & 0.7 & $-5.85^{+0.18}_{-0.25}$ & $-23.56^{+0.45}_{-0.72}$ & 0.0 \\
\hline
$2.5\pm0.1$ & 1 & 519 & $-3.48^{+0.01}_{-0.01}$ & $-19.46^{+0.03}_{-0.04}$ & $-0.9^{+0.1}_{-0.1}$ & 33.9 & $-3.86^{+0.05}_{-0.06}$ & $-20.65^{+0.11}_{-0.15}$ & 14.0 \\
$3.1\pm0.4$ & 5 & 1713 & $-3.60^{+0.01}_{-0.00}$ & $-19.77^{+0.02}_{-0.02}$ & $-1.1^{+0.1}_{-0.1}$ & 86.8 & $-3.59^{+0.03}_{-0.04}$ & $-20.50^{+0.04}_{-0.06}$ & 17.6 \\
$3.9\pm0.3$ & 2 & 169 & $-4.32^{+0.03}_{-0.02}$ & $-20.10^{+0.04}_{-0.11}$ & $-0.8^{+0.1}_{-0.1}$ & 3.3 & $-4.30^{+0.09}_{-0.08}$ & $-20.87^{+0.11}_{-0.13}$ & 2.4 \\
$4.7\pm0.2$ & 3 & 160 & $-4.44^{+0.02}_{-0.02}$ & $-20.73^{+0.03}_{-0.06}$ & $-0.9^{+0.1}_{-0.1}$ & 11.6 & $-4.62^{+0.05}_{-0.05}$ & $-21.73^{+0.08}_{-0.09}$ & 2.5 \\
$5.4\pm0.5$ & 4 & 167 & $-4.69^{+0.03}_{-0.03}$ & $-21.26^{+0.07}_{-0.07}$ & $-0.9^{+0.1}_{-0.1}$ & 8.5 & $-4.62^{+0.05}_{-0.06}$ & $-21.92^{+0.07}_{-0.09}$ & 0.0 \\
Full SC4K & 16 & 2857 & $-3.98^{+0.00}_{-0.01}$ & $-20.45^{+0.01}_{-0.04}$ & $-1.1^{+0.1}_{-0.1}$ & 191.3 & $-4.39^{+0.02}_{-0.02}$ & $-21.36^{+0.03}_{-0.04}$ & 14.6 \\
\hline
\multicolumn{10}{c|}{$\log_{10} (\rm L_{Ly\alpha}) \geq 43.0\,erg\,s^{-1}$} \\
\hline
$2.5\pm0.1$ & 1 & 47 & $-4.79^{+0.09}_{-0.20}$ & $-21.03^{+0.36}_{-1.98}$ & $-0.1^{+0.1}_{-0.1}$ & 1.2 & $-5.60^{+0.30}_{-0.31}$ & $-22.94^{+0.83}_{-1.03}$ & 0.8 \\
$3.1\pm0.4$ & 5 & 411 & $-4.43^{+0.02}_{-0.02}$ & $-20.32^{+0.05}_{-0.06}$ & $0.4^{+0.1}_{-0.1}$ & 13.5 & $-4.84^{+0.08}_{-0.09}$ & $-21.70^{+0.18}_{-0.25}$ & 5.5 \\
$3.9\pm0.3$ & 2 & 107 & $-4.70^{+0.03}_{-0.03}$ & $-20.40^{+0.07}_{-0.09}$ & $0.4^{+0.1}_{-0.1}$ & 5.1 & $-4.71^{+0.10}_{-0.11}$ & $-21.12^{+0.14}_{-0.20}$ & 0.5 \\
$4.7\pm0.2$ & 3 & 132 & $-4.54^{+0.02}_{-0.02}$ & $-20.75^{+0.04}_{-0.05}$ & $0.4^{+0.1}_{-0.1}$ & 8.8 & $-4.61^{+0.05}_{-0.06}$ & $-21.60^{+0.08}_{-0.10}$ & 1.4 \\
$5.4\pm0.5$ & 4 & 91 & $-4.88^{+0.04}_{-0.03}$ & $-21.33^{+0.08}_{-0.11}$ & $0.4^{+0.1}_{-0.1}$ & 5.3 & $-4.85^{+0.06}_{-0.07}$ & $-22.03^{+0.09}_{-0.11}$ & 0.1 \\
Full SC4K & 16 & 789 & $-4.65^{+0.02}_{-0.01}$ & $-20.79^{+0.03}_{-0.04}$ & $0.4^{+0.1}_{-0.1}$ & 27.8 & $-5.03^{+0.03}_{-0.04}$ & $-21.95^{+0.07}_{-0.08}$ & 7.7 \\
\hline
\end{tabular}
\end{table*}

\begin{table*}
\setlength{\tabcolsep}{5pt}
\caption{Best-fit Schechter parameters for the SMF of LAEs from $z=2$ to $z=6$, for each of the individual selection filters and for different redshift bins (see \S\ref{subsec:binning}). The number of sources provided here is the number of sources included in the stellar mass functions, i.e. non-AGN LAEs with available SEDs and with completeness corrections $>30\%$. We provide best fits for the two cases considered in this study: fit to the full stellar mass range (blue in Fig. \ref{fig:grid_smf}) and fit to the bins more massive than the number density peak (purple in Fig. \ref{fig:grid_smf}). We provide the best set of parameters ($\alpha$, M$_\star^*$ and $\Phi^*$) which minimise $\chi^2_{\rm red}$, with $\alpha$ being fixed for the latter case as it cannot be directly constrained. When $\chi^2_{\rm red}$ is very large, the errors should be interpreted with caution as the best parameters found still do not provide a good model. Additionally, M$_\star^*$ is also fixed for the individual filters with less than three luminosity bins (although we perturb these parameters when exploring the uncertainties of the bins/fits, see \S\ref{subsec:perturb_fits}). For the redshift bins we also show the Schechter parameters when applying a $\log_{10} (\rm L_{Ly\alpha}/erg\,s^{-1}) \geq 43.0$ cut.} \label{tab:schechter_params_smf}
\begin{tabular}{ccc | cccc | ccc}
\hline
  & & & \multicolumn{4}{c}{Full M$_\star$ range}  |& \multicolumn{3}{c}{M$_\star$ above the peak}\\ 
   & & & \multicolumn{4}{c}{}  & \multicolumn{3}{c}{($\alpha_{\rm fix}=-1.3$)}\\
\hline
Redshift & \# Filters & \# Sources & $\log_{10}\Phi^*$ & M$_\star^*$ & $\alpha^*$ & $\chi^2_{\rm red}$ & $\log_{10}\Phi^*$ & M$_\star^*$ & $\chi^2_{\rm red}$\\
&  &  & (Mpc$^{-3}$) & (AB) & & &(Mpc$^{-3}$)&(AB)& \\
\hline
$2.2\pm0.1$ & 1 & 129 & $-3.94^{+0.08}_{-0.07}$ & $10.41^{+0.06}_{-0.06}$ & $-0.7^{+0.1}_{-0.1}$ & 7.8 & $-4.44^{+0.04}_{-0.03}$ & $10.69^{+0.05}_{-0.05}$ & 7.7 \\
$2.5\pm0.1$ & 1 & 519 & $-4.14^{+0.01}_{-0.01}$ & $10.40^{+0.02}_{-0.02}$ & $-0.9^{+0.1}_{-0.1}$ & 76.2 & $-4.66^{+0.02}_{-0.02}$ & $10.67^{+0.04}_{-0.03}$ & 13.4 \\
$2.8\pm0.1$ & 1 & 139 & $-5.90^{+0.20}_{-0.40}$ & $11.41^{+0.79}_{-0.19}$ & $-1.5^{+0.1}_{-0.1}$ & 17.6 & $-5.33^{+0.04}_{-0.05}$ & $11.10^{+0.12}_{-0.08}$ & 14.1 \\
$3.0\pm0.1$ & 1 & 565 & $-4.79^{+0.02}_{-0.03}$ & $10.73^{+0.05}_{-0.03}$ & $-1.4^{+0.1}_{-0.1}$ & 67.9 & $-4.57^{+0.02}_{-0.01}$ & $10.61^{+0.02}_{-0.03}$ & 15.4 \\
$3.2\pm0.1$ & 1 & 31 & $-4.25^{+0.09}_{-0.11}$ & $10.05^{+0.15}_{-0.09}$ & $-0.3^{+0.1}_{-0.1}$ & 3.6 & $-4.68^{+0.07}_{-0.08}$ & $10.31^{+0.12}_{-0.07}$ & 3.1 \\
$3.2\pm0.1$ & 1 & 413 & $-4.03^{+0.01}_{-0.05}$ & $10.33^{+0.03}_{-0.02}$ & $-0.7^{+0.1}_{-0.1}$ & 39.4 & $-4.62^{+0.02}_{-0.01}$ & $10.66^{+0.02}_{-0.03}$ & 12.8 \\
$3.3\pm0.1$ & 1 & 565 & $-4.19^{+0.01}_{-0.01}$ & $10.76^{+0.03}_{-0.03}$ & $-0.9^{+0.1}_{-0.1}$ & 58.0 & $-4.60^{+0.02}_{-0.02}$ & $10.93^{+0.04}_{-0.03}$ & 29.7 \\
$3.7\pm0.1$ & 1 & 53 & $-5.50^{+0.16}_{-0.24}$ & $10.72^{+0.19}_{-0.10}$ & $-1.3^{+0.1}_{-0.1}$ & 28.6 & $-5.40^{+0.04}_{-0.04}$ & $10.66^{+0.07}_{-0.06}$ & 9.9 \\
$4.1\pm0.1$ & 1 & 116 & $-4.86^{+0.03}_{-0.09}$ & $10.37^{+0.09}_{-0.04}$ & $-0.8^{+0.1}_{-0.1}$ & 19.0 & $-5.38^{+0.04}_{-0.04}$ & $10.63^{+0.07}_{-0.05}$ & 8.4 \\
$4.6\pm0.1$ & 1 & 69 & $-5.65^{+0.11}_{-0.34}$ & $10.74^{+0.40}_{-0.10}$ & $-1.3^{+0.1}_{-0.1}$ & 3.7 & $-5.64^{+0.05}_{-0.06}$ & $10.74^{+0.11}_{-0.08}$ & 0.4 \\
$4.8\pm0.1$ & 1 & 50 & $-5.04^{+0.07}_{-0.13}$ & $10.49^{+0.10}_{-0.05}$ & $-0.9^{+0.1}_{-0.1}$ & 15.8 & $-5.47^{+0.04}_{-0.05}$ & $10.73^{+0.10}_{-0.06}$ & 1.9 \\
$4.8\pm0.1$ & 1 & 41 & $-4.53^{+0.07}_{-0.08}$ & $10.51^{+0.10}_{-0.06}$ & $-0.7^{+0.1}_{-0.1}$ & 0.3 & $-4.79^{+0.05}_{-0.04}$ & $10.68\,{\rm (fix)}$ & 0.1 \\
$5.0\pm0.1$ & 1 & 29 & $-4.96^{+0.07}_{-0.08}$ & $10.32^{+0.07}_{-0.05}$ & $-0.4^{+0.1}_{-0.1}$ & 4.6 & $-5.52^{+0.05}_{-0.05}$ & $10.68^{+0.08}_{-0.06}$ & 0.2 \\
$5.2\pm0.1$ & 1 & 17 & $-5.10^{+0.04}_{-0.06}$ & $9.83^{+0.11}_{-0.06}$ & $0.0^{+0.1}_{-0.1}$ & 3.4 & $-5.75^{+0.04}_{-0.04}$ & $10.68\,{\rm (fix)}$ & 0.2 \\
$5.7\pm0.1$ & 1 & 107 & $-4.80^{+0.16}_{-0.10}$ & $11.38^{+0.11}_{-0.09}$ & $-1.0^{+0.1}_{-0.1}$ & 11.8 & $-5.18^{+0.04}_{-0.05}$ & $11.54^{+0.12}_{-0.07}$ & 7.3 \\
$5.8\pm0.1$ & 1 & 14 & $-5.35^{+0.11}_{-0.13}$ & $10.34^{+0.14}_{-0.09}$ & $-0.5^{+0.1}_{-0.1}$ & 1.4 & $-5.86^{+0.08}_{-0.10}$ & $10.66^{+0.17}_{-0.09}$ & 0.1 \\
\hline
$2.5\pm0.1$ & 1 & 519 & $-4.14^{+0.01}_{-0.01}$ & $10.40^{+0.02}_{-0.02}$ & $-0.9^{+0.1}_{-0.1}$ & 76.2 & $-4.66^{+0.02}_{-0.02}$ & $10.67^{+0.04}_{-0.03}$ & 13.4 \\
$3.1\pm0.4$ & 5 & 1713 & $-4.57^{+0.01}_{-0.01}$ & $10.86^{+0.02}_{-0.02}$ & $-1.1^{+0.1}_{-0.1}$ & 193.2 & $-4.90^{+0.01}_{-0.01}$ & $11.02^{+0.03}_{-0.03}$ & 55.2 \\
$3.9\pm0.3$ & 2 & 169 & $-4.89^{+0.07}_{-0.02}$ & $10.40^{+0.03}_{-0.05}$ & $-0.8^{+0.1}_{-0.1}$ & 47.8 & $-5.39^{+0.02}_{-0.03}$ & $10.64^{+0.05}_{-0.03}$ & 18.9 \\
$4.7\pm0.2$ & 3 & 160 & $-5.07^{+0.06}_{-0.02}$ & $10.53^{+0.03}_{-0.05}$ & $-0.9^{+0.1}_{-0.1}$ & 25.7 & $-5.52^{+0.02}_{-0.04}$ & $10.81^{+0.07}_{-0.05}$ & 0.0 \\
$5.4\pm0.5$ & 4 & 167 & $-5.31^{+0.07}_{-0.09}$ & $11.19^{+0.07}_{-0.05}$ & $-0.9^{+0.1}_{-0.1}$ & 26.9 & $-5.78^{+0.03}_{-0.04}$ & $11.37^{+0.07}_{-0.05}$ & 14.2 \\
Full SC4K & 16 & 2857 & $-4.93^{+0.00}_{-0.01}$ & $10.94^{+0.02}_{-0.01}$ & $-1.1^{+0.1}_{-0.1}$ & 372.4 & $-5.19^{+0.01}_{-0.00}$ & $11.05^{+0.01}_{-0.01}$ & 95.2 \\
\hline
\multicolumn{10}{c|}{$\log_{10} (\rm L_{Ly\alpha}) \geq 43.0\,erg\,s^{-1}$} \\
\hline
$2.5\pm0.1$ & 1 & 47 & $-4.79^{+0.09}_{-0.20}$ & $-21.03^{+0.36}_{-1.98}$ & $-0.6^{+0.1}_{-0.1}$ & 1.2 & $-5.60^{+0.30}_{-0.31}$ & $-22.94^{+0.83}_{-1.03}$ & 0.8 \\
$3.1\pm0.4$ & 5 & 411 & $-4.43^{+0.02}_{-0.02}$ & $-20.32^{+0.05}_{-0.06}$ & $-1.2^{+0.1}_{-0.1}$ & 13.5 & $-4.84^{+0.08}_{-0.09}$ & $-21.70^{+0.18}_{-0.25}$ & 5.5 \\
$3.9\pm0.3$ & 2 & 107 & $-4.70^{+0.03}_{-0.03}$ & $-20.40^{+0.07}_{-0.09}$ & $-0.8^{+0.1}_{-0.1}$ & 5.1 & $-4.71^{+0.10}_{-0.11}$ & $-21.12^{+0.14}_{-0.20}$ & 0.5 \\
$4.7\pm0.2$ & 3 & 132 & $-4.54^{+0.02}_{-0.02}$ & $-20.75^{+0.04}_{-0.05}$ & $-1.0^{+0.1}_{-0.1}$ & 8.8 & $-4.61^{+0.05}_{-0.06}$ & $-21.60^{+0.08}_{-0.10}$ & 1.4 \\
$5.4\pm0.5$ & 4 & 91 & $-4.88^{+0.04}_{-0.03}$ & $-21.33^{+0.08}_{-0.11}$ & $-0.9^{+0.1}_{-0.1}$ & 5.3 & $-4.85^{+0.06}_{-0.07}$ & $-22.03^{+0.09}_{-0.11}$ & 0.1 \\
Full SC4K & 16 & 789 & $-4.65^{+0.02}_{-0.01}$ & $-20.79^{+0.03}_{-0.04}$ & $-1.1^{+0.1}_{-0.1}$ & 27.8 & $-5.03^{+0.03}_{-0.04}$ & $-21.95^{+0.07}_{-0.08}$ & 7.7 \\
\hline\end{tabular}
\end{table*}

\subsection{Fitting the UV luminosity function} \label{subsec:muv_lf}

In order to compare our results with previous studies, we adopt the common parameterisation of \cite{Schechter1976} function, which consists of a power-law with a slope $\alpha$ for faint luminosities and a declining exponential for brighter luminosities. The transition between the two regimes is given by a characteristic luminosity (L$^*$) and a characteristic number density ($\Phi^*$). The Schechter equation has the following form:

\begin{eqnarray}
\rm \Phi(L)=\frac{\Phi^*}{L^*}\left(\frac{L}{L^*} \right)^\alpha \exp\left(-\frac{L}{L^*}\right),
\end{eqnarray} \label{eq:schechter}

Equation \ref{eq:schechter} can be rewritten for absolute magnitudes by using the substitution $\Phi(L)\mathrm{d}L=\Phi($M$_{\rm UV})\mathrm{d}$M$_{\rm UV}$:

\begin{eqnarray}
\Phi(\rm M_{\rm UV})=\frac{\ln 10}{2.5}\,\Phi^*\,10^{-0.4(\alpha+1)\Delta M_{\rm UV}}\exp\left(-10^{-0.4\Delta M_{\rm UV}}\right),
\end{eqnarray}

\noindent where $\rm \Delta M_{\rm UV}=M_{\rm UV}-M_{\rm UV}^*$. 

The observed UV luminosity distribution of LAEs shows the same behaviour at all redshifts: there is a peak number density at an intermediate UV luminosity, with a subsequent decline in number density for both brighter and fainter UV luminosities (see Fig. \ref{fig:grid_muv_lf}). While such a distribution does not resemble the Schechter function with a steep faint end which is typically measured in LBG samples \citep[e.g.][]{Bouwens2015,Finkelstein2015}, we argue that such observed distribution of UV luminosities can be expected for a sample which is selected by Ly$\alpha$ line flux above some threshold (corresponding to a vertical cut in Fig. \ref{fig:lya_vs_others}), causing an incomplete sampling of M$_{\rm UV}$. This incomplete sampling is most significant at the faint UV luminosities, which is shown in Fig. \ref{fig:lf_full} (right panel) where an increasing L$_{\rm Ly\alpha}$ limit will cause a preferential decline of number densities at faint UV luminosities and hence the observed turn-over. Thus, in order to conduct a detailed analysis of the UV luminosity distribution of LAEs, we explore two separate scenarios:

\begin{itemize}
\item fit to the full UV luminosity range (blue in Fig. \ref{fig:grid_muv_lf}): the entire observable UV luminosity range is considered, including the turn-over at faint UV luminosities. While the low number densities at faint UV luminosities may be driven by our L$_{\rm Ly\alpha}$ limits, this approach provides the best-fit to the directly observed number densities.
\item fit to the UV luminosity range brighter than the number density peak (purple in Fig. \ref{fig:grid_muv_lf}): the bins fainter than the number density peak (dominated by an incomplete sampling) are thus not included in the fitting, and the faint UV luminosity regime becomes unconstrained. The peak in number density is different for different filters (see Fig. \ref{fig:grid_muv_lf}) and different L$_{\rm Ly\alpha}$ limits (see Fig. \ref{fig:lf_full}, right panel). With the simple assumptions of a steep faint end slope (as measured in UV luminosity-selected samples) and by not including the bins below of the turn-over (which are heavily dominated by our L$_{\rm Ly\alpha}$ limits), we obtain a proxy of the full distribution of LAEs.
\end{itemize}

We provide the Schechter parameters of the best-fits to both cases in Table \ref{tab:schechter_params_uv_lf}. For the fit to the full UV luminosity range, we find the set of parameters ($\alpha$, M$_{\rm UV}^*$, $\Phi^*$) which minimises the reduced $\chi^2$ ($\chi^2_{\rm red}$) in log-space. Alternatively, fitting can also be performed in linear-space, where $\chi^2$ is less sensitive to bins with low number densities. A fit in log-space thus tends to favour slightly brighter characteristic luminosities which provide a better fit to the very bright luminosities. We find that the observed distribution is best fit by shallow faint end slopes ($\alpha>-1$), which are able to represent the turn-over at the faintest luminosities, with $\alpha$ even being positive at some redshift ranges.

When constraining only the UV luminosities brighter than the number density peak, we are not able to directly constrain the $\alpha$ slope of the power-law, and thus fix $\alpha$ to -1.5 \citep[similar to the UV LF of LAEs from e.g.][]{Ouchi2008}, but we still perturb this parameter to quantify uncertainties (see \S\ref{subsec:perturb_fits}). Here, we make the assumption that $\alpha$ does not evolve with redshift, which is a necessary caveat due to not being able to directly constrain it. We measure the UV LF of LAEs selected in each MB or NB by determining the pair (M$_{\rm UV}^*$, $\Phi^*$) which minimises $\chi^2_{\rm red}$ in log-space of the M$_{\rm UV}$ luminosity bins with associated Poissonian error bars. In Fig. \ref{fig:grid_muv_lf}, we show the luminosity bins and luminosity functions of LAEs from the 16 selection filters. For the filters with only two luminosity bins brighter than the number density peak, we can only fit one free parameter, so we fix M$_{\rm UV}^*$ to a similar nearby filter (NB501 uses M$_{\rm UV, IA505}^*=-20.37$ and IA738+IA767 use M$_{\rm UV, IA709}^*=-21.22$). We provide the Schechter parameters of the best fits in Table \ref{tab:schechter_params_uv_lf}.

\subsection{Fitting the stellar mass function} \label{subsec:smf}

Following a similar logic to what was done in \S\ref{subsec:muv_lf}, Equation \ref{eq:schechter} can be rewritten in $\rm\log M$ space:

\begin{eqnarray}
\Phi(\rm M_\star)=\ln 10\,\Phi^*\,10^{(\alpha+1)\Delta M}\exp\left(-10^{\Delta M}\right),
\end{eqnarray}

\noindent where $\rm \Delta M=\log_{10} M_\star-\log_{10} M_\star^*$.  At $z<1$, a double Schechter function has been commonly used \citep[see e.g.][]{Pozzetti2010,Ilbert2013}, with two $\alpha$ and two $\Phi^*$, which are capable of reproducing a bimodal population, which includes quiescent galaxies. In this work, we restrain ourselves to a single Schechter as the quiescent population should not contribute to our Ly$\alpha$-selected sample, particularly at the redshift range that we probe.

Similarly to the observed UV LF, the observed number density distribution of the stellar mass peaks at an intermediate stellar mass, and declines for both lower and higher stellar masses (see Fig. \ref{fig:grid_smf}). While a Schechter distribution with a steep slope could be expected for a mass-selected sample, as our LAEs are selected by being above some Ly$\alpha$ line flux (corresponding to a vertical cut in Fig. \ref{fig:lya_vs_others}) determined by observational constraints, there is a turn-over at low stellar masses. The preferential decline of low stellar masses with increasing Ly$\alpha$ line flux is shown in Fig. \ref{fig:smf_full} (right panel), and we further discuss how to interpret the shape of the SMF in \S\ref{sec:interpret_LF}.

Following the same reasons listed for the UV LF, we conduct our fitting procedure in two stellar mass ranges: full stellar mass range (blue in Fig. \ref{fig:grid_smf}) and stellar mass range above the turn-over, with an assumption of the $\alpha$ slope (blue in Fig. \ref{fig:grid_smf}). The former provides a fit to the directly observed number densities and the later provides a proxy SMF of the full distribution of LAEs. We provide the best Schechter fits to both cases in Table \ref{tab:schechter_params_smf}. For the fit of the full stellar mass range, we find the set of parameters ($\alpha$, M$_\star^*$, $\Phi^*$) which minimises $\chi^2_{\rm red}$ in log-space. The observed distribution with a turn-over for the faintest luminosities, results in shallow faint end slopes ($\alpha>-1$).

When constraining only the stellar masses bigger than the number density peak, we are not able to directly constrain the $\alpha$ slope of the power-law. We fix $\alpha$ to -1.3, but we vary all parameters, including $\alpha$ in \S\ref{subsec:perturb_fits}. Similarly to the UV LF, we introduce the caveat the $\alpha$ does not evolve with redshift, which is a necessary assumption due to us not being able to directly constrain it. In Fig. \ref{fig:grid_smf}, we show the stellar mass bins and SMFs of LAEs from the 16 selection filters. For the filters with only two stellar mass bins, we can only fit one free parameter, so we fix M$_\star^*$ to a similar nearby filter (NB711+IA767 use M$_{\rm \star, IA738}^*=10^{10.68}$\,M$_\odot$). We provide the Schechter parameters of in best fits in Table \ref{tab:schechter_params_smf}.

\subsection{Perturbing the luminosity and mass functions} \label{subsec:perturb_fits}
We explore the uncertainties in our UV LFs and SMFs by perturbing the luminosity or mass bins within their Poissonian error bars. For each iteration, we perturb each bin within their error bars (assuming a normal probability distribution function centred at each bin and with FWHM equal to the error) and determine the value for the current realisation. We compute the best Schechter fit to the bins of the current realisation and iterate the process 1000 times. We obtain the 16th and 84th percentile of all fits, which we plot as contours in all figures. For each iteration, we also perturb the fixed Schechter parameters ($\alpha$ for all redshifts and M$_\star^*$ or M$_{\rm UV}^*$ for the filters with only two bins) by picking a random value in a $\pm0.2$ dex range centred in the fixed values \citep[same method as][]{Sobral2018}.

%
%
\begin{figure*}
  \centering
\begin{tabular}{ll}
  \centering
  \includegraphics[width=0.487\textwidth]{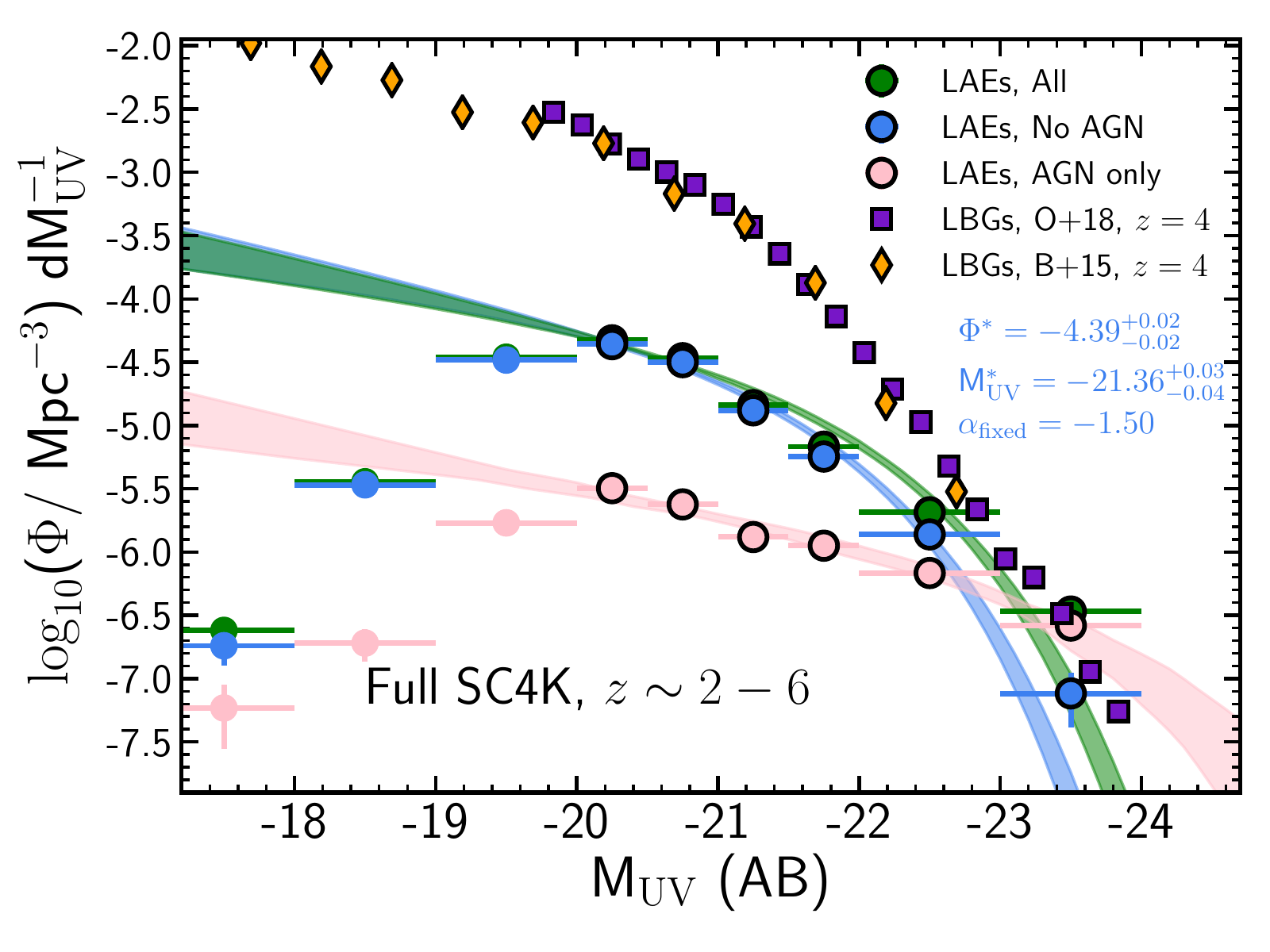}
  &
  \includegraphics[width=0.487\textwidth]{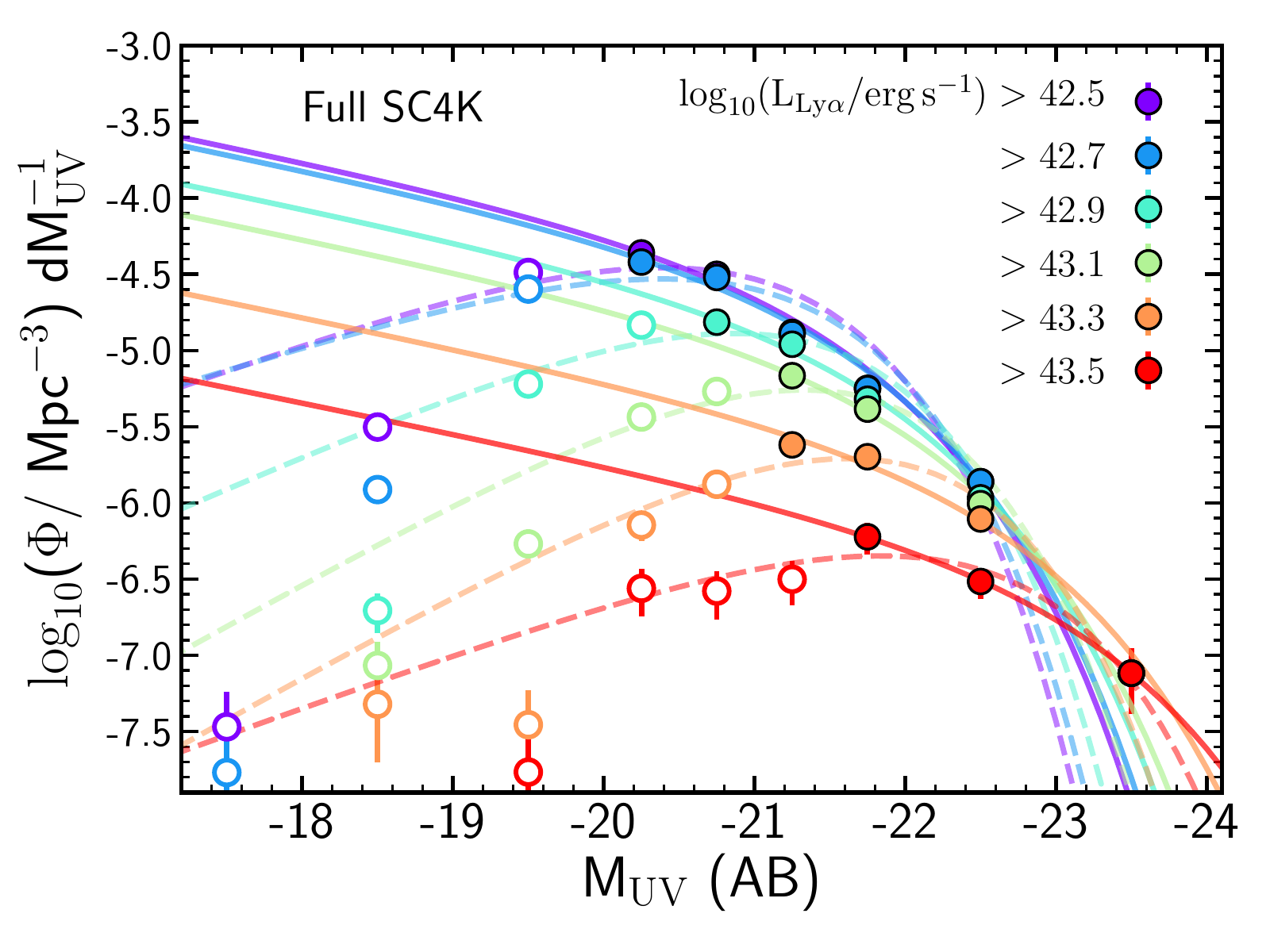}  
\end{tabular}
  \caption{{\it Left:} UV LF of the full SC4K sample of LAEs: including AGN (green), no AGN (blue; what we use throughout this work) and AGN only (pink). AGN LAEs dominate the bright end ($-24<$\,M$_{\rm UV}<-23$) of the UV LF of LAEs. The contours are the 16th and 84th percentiles, which we obtain by perturbing the bins within their Poissonian error bars and iterating the fitting 1000 times (see \S\ref{subsec:perturb_fits}). For reference, we show UV LFs of LBGs at $z\sim4$, from \citet{Bouwens2015} (orange diamonds) and \citet{Ono2018} (purple squares). The number density of M$_{\rm UV}=-20$ LAEs is $\sim1.5$ dex lower than LBGs, but they converge to the same number densities at M$_{\rm UV}<-23$. {\it Right:} UV LF of the full SC4K sample at different L$_{\rm Ly\alpha}$ cuts. We show the best Schechter fits to the full UV luminosity range as dashed lines, and to the number densities above the turn-over as filled lines (see \S\ref{subsec:muv_lf}). The increasing L$_{\rm Ly\alpha}$ cuts reduce the number densities, predominantly for fainter M$_{\rm UV}$, which can be linked with L$_{\rm Ly\alpha}$ and M$_{\rm UV}$ being typically correlated (see Fig. \ref{fig:lya_vs_others}, left panel). However, note that the UV LF of more luminous LAEs yields a declining $\Phi^*$ but a brightening in M$_{\rm UV}^*$.}
\label{fig:lf_full}
\end{figure*}

\subsection{Obtaining UV and stellar mass densities} \label{subsec:methods_densities}

We integrate UV LFs and SMFs to obtain the luminosity density ($\rho_{\rm UV}$) and the stellar mass density ($\rho_{\rm M}$), respectively. In order to fully take into account the uncertainties in our luminosities/stellar masses, we perturb our measurements within their errors and fit and integrate each of the 1000 realisations (see \S\ref{subsec:perturb_fits}). The computed $\rho_{\rm UV}$ and $\rho_{\rm M}$ are the median of all integrals, with the errors being the 16th and 84th percentile of the distribution of all realisations. To obtain $\rho_{\rm UV}$, we compute the integral of the UV LFs in the range $-23 < {\rm M_{UV}} < -17$ \citep[similar to e.g.][]{Finkelstein2015,Bouwens2015}. To obtain $\rho_{\rm M}$, we compute the integral of the SMFs in the range $10^{8-13}$\,M$_\odot$ \citep[similar to e.g.][]{Davidzon2017}. All $\rho_{\rm M}$ measurements in this study assume a Chabrier IMF, and values from the literature are converted to a Chabrier IMF if another IMF was used. 

\section{Results and Discussion} \label{sec:results}

\subsection{Interpreting the observed UV LF and SMF} \label{sec:interpret_LF}
As detailed in the previous sections, the observed distribution of both the UV LF and SMF of LAEs has a turn-over at the faintest UV luminosities and smallest stellar masses, respectively (see Fig. \ref{fig:grid_muv_lf} and \ref{fig:grid_smf}). While such a turn-over has not been observed in UV-selected or mass-selected samples, it is an expected distribution of a Ly$\alpha$-selected sample, where Ly$\alpha$ correlates with both M$_{\rm UV}$ and M$_\star$ but with significant scatter (see Fig. \ref{fig:lya_vs_others}), as there will be incomplete sampling of M$_{\rm UV}$ and M$_\star$, particularly at the faint UV luminosity and low stellar mass regimes.  As shown in Fig. \ref{fig:lf_full} (right panel), an increasing L$_{\rm Ly\alpha}$ cut will preferentially decrease the number densities of the faintest UV luminosities, creating the turn-over which is a consequence of selection and not an intrinsic property of the UV LF of LAEs. A similar dependence is measured for the SMF in Fig. \ref{fig:smf_full} (right panel), albeit the dependence is not as strong. We make the assumption that the incomplete sampling will introduce only small contributions above the turn-over, which is supported by our measurements (Fig. \ref{fig:lf_full}, right panel): when extending the luminosity cut from $\log_{10} (\rm L_{Ly\alpha}/erg\,s^{-1})=42.5$ to 42.7 (and even further into 42.9 and 43.1), the number densities always have a very significant drop below the turn-over but remain roughly constant above it. By only fitting the regime above the turn-over and by fixing $\alpha$ as a steep slope, we are able to measure a distribution which is not dominated by incomplete sampling, and compute a proxy for the full UV LFs and SMFs.

We provide in Table \ref{tab:schechter_params_uv_lf} and \ref{tab:schechter_params_smf} the best Schechter parameters of the distribution of 1) the full UV luminosity (or stellar mass) ranges (see the blue contours in Fig. \ref{fig:grid_muv_lf} and \ref{fig:grid_smf}) and 2) the UV luminosity (or stellar mass) range above the turn-over, with a fixed steep $\alpha$ slope (see the purple contours in Fig. \ref{fig:grid_muv_lf} and \ref{fig:grid_smf}). As we aim to understand the full LAE population, in the analysis conducted in the following sections we use the second fitting procedure, which gives a proxy of the full distribution of LAEs. We note nonetheless that the L$_{\rm Ly\alpha}$ limits can have some influence on the number densities even above the turn-over, so when probing redshift evolution we extend the analysis to always use the same L$_{\rm Ly\alpha}$ cut and ensure the samples are comparable (see discussion in \S\ref{sec:muv_vary}).

%
%
\begin{figure*}
\begin{tabular}{ll}
  \centering
  \includegraphics[width=0.487\textwidth]{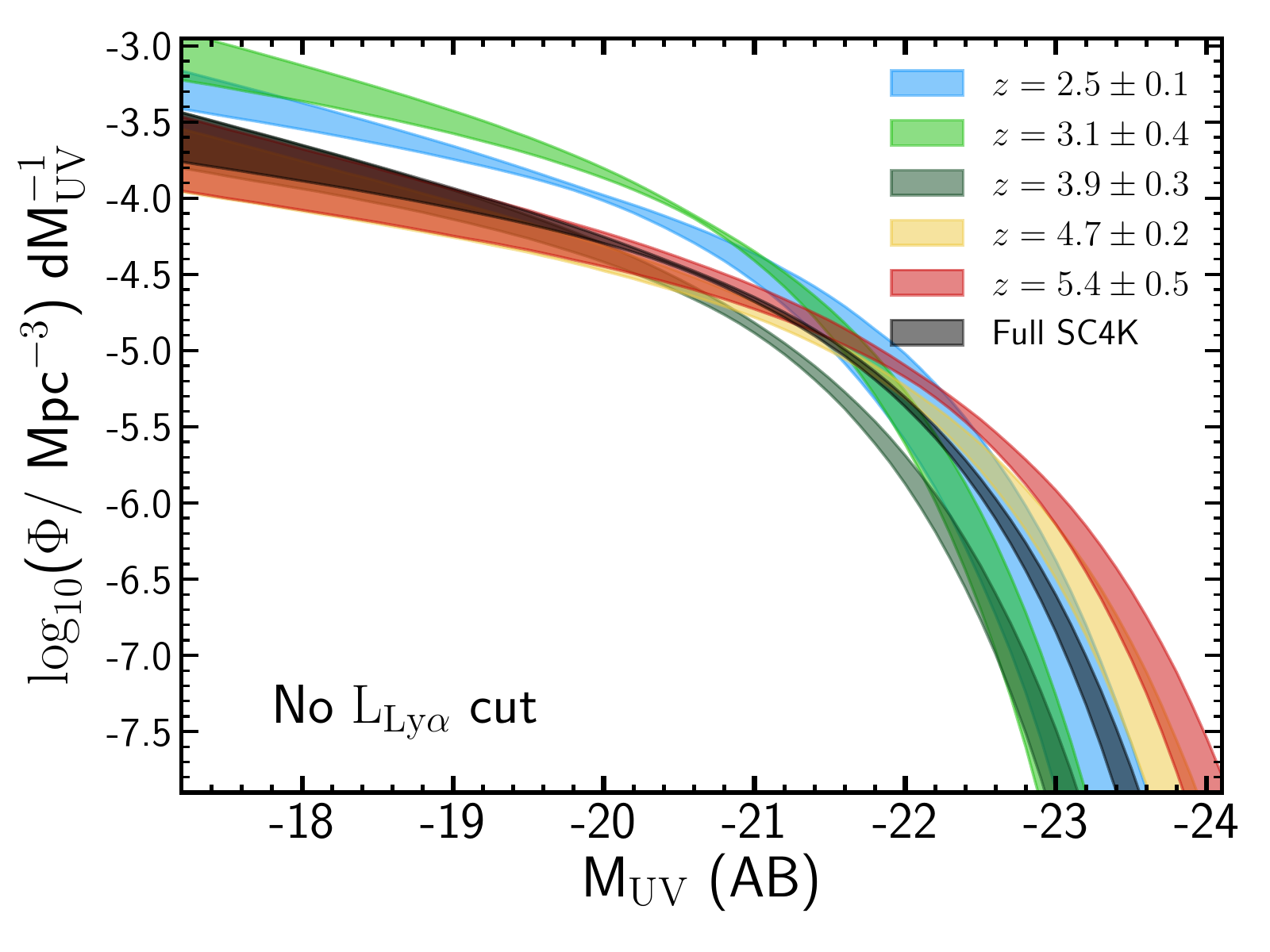}
  &
  \includegraphics[width=0.487\textwidth]{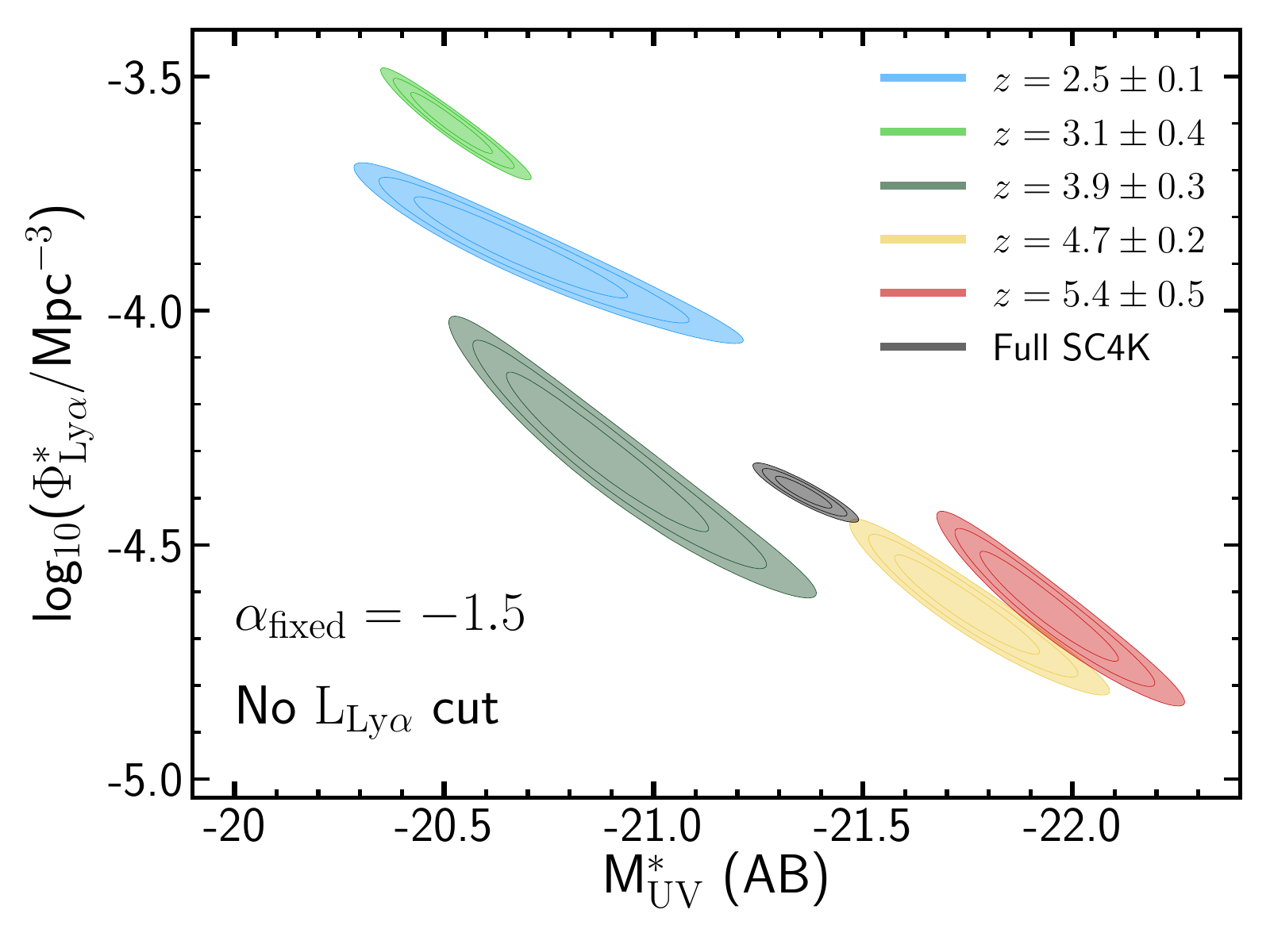}  
\end{tabular}
  \caption{{\it Left:} Evolution of the UV LF with redshift, with no L$_{\rm Ly\alpha}$ cut. The shaded contours are the 16th and 84th percentiles of all iterations obtained by perturbing the luminosity bins (see \S\ref{subsec:perturb_fits}) {\it Right:} $\Phi^*-$M$_{\rm UV}^*$ 1$\sigma$, 2$\sigma$ and 3$\sigma$ contours. We observe an M$_{\rm UV}^*$ increase from $\sim-20.5$ at $z\sim2.5$ to $\sim-22$ at $z\sim5-6$, and a $\log_{10}(\Phi^*/$Mpc$^{-3})$ decrease from $\sim-3.5$ to $\sim-4.5$ for the same redshifts.}
  \label{fig:multiple_lf_no_cut}
\end{figure*}

%
%
\begin{figure*}
\begin{tabular}{ll}
  \centering
  \includegraphics[width=0.487\textwidth]{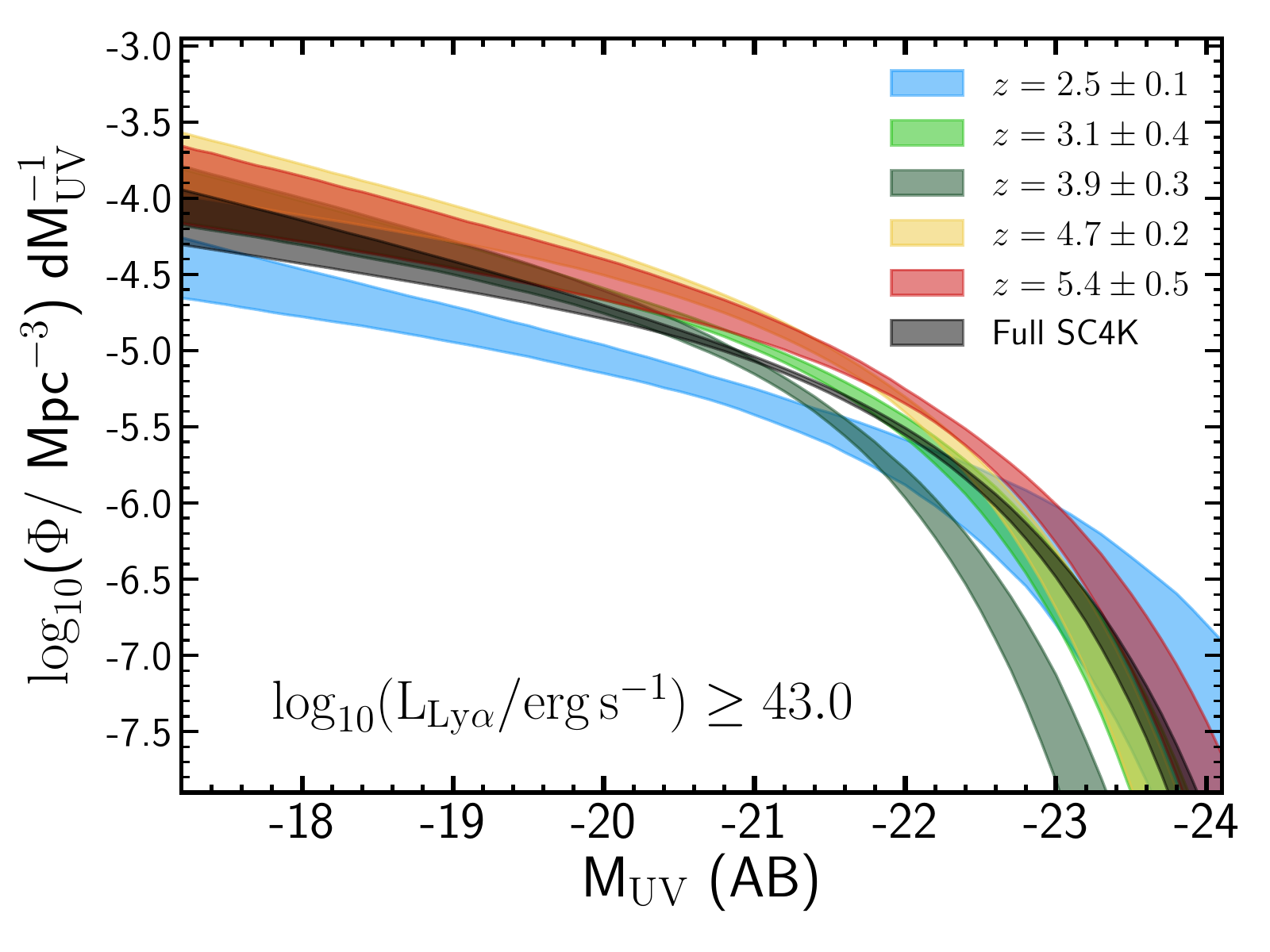}
  &
  \includegraphics[width=0.487\textwidth]{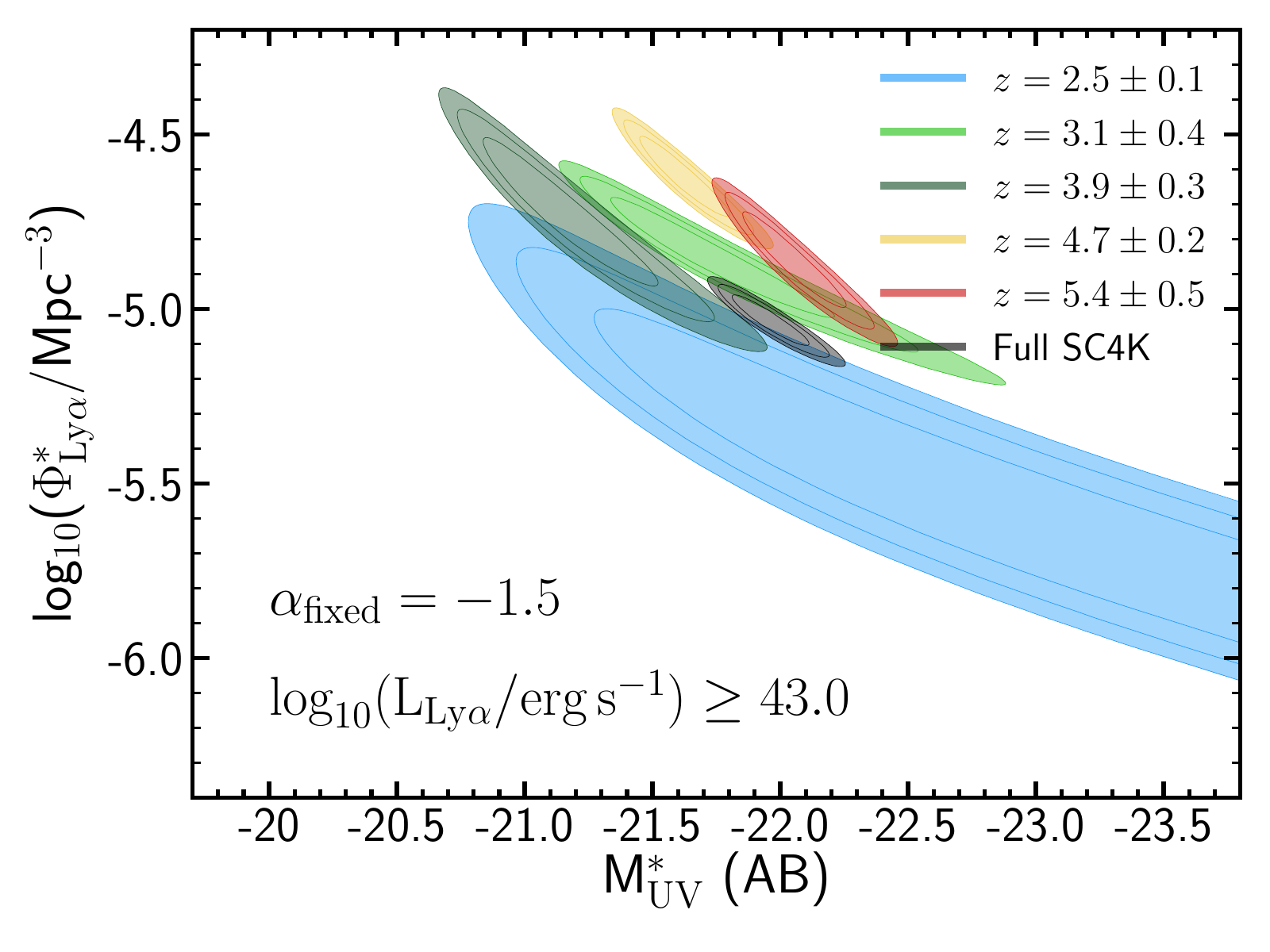}  
\end{tabular}
  \caption{{\it Left:} Evolution of the UV LF with redshift, with a luminosity cut of $\log_{10} (\rm L_{Ly\alpha}/erg\,s^{-1}) \geq 43.0$. {\it Right:} $\Phi^*-$M$_{\rm UV}^*$ 1$\sigma$, 2$\sigma$ and 3$\sigma$ contours. With a uniform cut for the entire sample, we note no clear evolutionary trend in M$_{\rm UV}^*$, while $\log_{10}(\Phi^*/$Mpc$^{-3})$ remains roughly constant at -4.7 at $z\sim3-6$.}
  \label{fig:multiple_lf_cut}
\end{figure*}

\subsection{The global UV LF of LAEs at $z\sim2-6$}  \label{sec:muv_lf_evo}

We start by measuring the UV LF of the full sample of SC4K LAEs, exploring a large volume of $\sim10^8$ Mpc$^3$ at $z\sim2-6$. With our large sample of $\sim4000$ LAEs, we are capable of probing extremely bright UV luminosities, down to M$_{\rm UV}=-24$, which even in UV-continuum searches has typically only been reached in very wide area ground-based surveys \citep[e.g.][]{Bowler2017,Ono2018}. Additionally, we have a statistically robust sample up to M$_{\rm UV}=-20$, providing a robust probe in a range of 4 dex in M$_{\rm UV}$, with individual LAEs as faint as M$_{\rm UV}=-17$. 

We show in Fig. \ref{fig:lf_full} (left panel) the UV LF for three subsets of SC4K LAEs: 1) All LAEs; 2) All LAEs after removing AGN (this is the subset we use throughout this paper; see \S\ref{sec:sample_AGNs}); 3) AGN LAEs only. We show the best Schechter fits to each case as 1$\sigma$ contours, which we obtain by perturbing the luminosity bins within their Poissonian errors and fitting 1000 realisations of the perturbed bins (see \S\ref{subsec:perturb_fits}). We find that the UV LF of all LAEs resembles a Schechter distribution, although there is an excess at M$_{\rm UV}<-23$, where the UV LF starts deviating from a Schechter function. A single power-law with best-fit $\log_{10}(\Phi/$Mpc$^{-3})=0.71^{+0.01}_{-0.01}$M$_{\rm UV}+10.26^{+0.11}_{-0.11}$ is also a very good fit ($\chi^2_{\rm reduced}=4.01$). When excluding AGNs, the number density significantly drops by 0.7 dex at the bright end ($-24<$\,M$_{\rm UV}<-23$), and the LF becomes steeper, with the single power law, with best-fit $\log_{10}(\Phi/$Mpc$^{-3})=0.91^{+0.01}_{-0.01}$M$_{\rm UV}+14.49^{+0.12}_{-0.11}$, becoming less preferable ($\chi^2_{\rm reduced}=60.63$). We observe that AGN LAEs clearly dominate the bright end ($-24<$\,M$_{\rm UV}<-23$) of the UV LF, with only minor contributions to the faint end ($-22<$\,M$_{\rm UV}<-20$). This trend is qualitatively similar to the one found by \cite{Sobral2018} for the Ly$\alpha$ LF of LAEs. Such a similar behaviour between the UV LF and Ly$\alpha$ LF is a consequence of L$_{\rm Ly\alpha}$ and M$_{\rm UV}$ being typically correlated (see Fig. \ref{fig:lya_vs_others}, left panel), although the complicated radiative transfer physics behind Ly$\alpha$ emission should be noted.

\subsection{UV LF with varying {\rm L$_{\rm Ly\alpha}$} cuts} \label{sec:muv_vary}

Due to an increasing luminosity distance with redshift, we are only capable of reaching the faintest Ly$\alpha$ luminosities (down to $10^{42.5}$\,erg\,s$^{-1}$) at $z\sim2.5$, or at higher redshifts with NBs. We aim to ensure that when comparing UV LFs at different redshifts, results are not driven by differences in depth. As such, we need to estimate how different Ly$\alpha$ luminosity limits affect the UV LF of LAEs. We show in Fig. \ref{fig:lf_full} (right panel) the UV LF of the full SC4K sample with varying ${\rm L_{Ly\alpha}}$ cuts, from $10^{42.5}$ to $10^{43.5}$\,erg\,s$^{-1}$. As expected from the dependence of M$_{\rm UV}$ and L$_{\rm Ly\alpha}$, an increasing ${\rm L_{Ly\alpha}}$ cut predominantly decreases the number densities of fainter M$_{\rm UV}$ LAEs. For the full SC4K sample, between $10^{42.5}$ and $10^{43.3}$\,erg\,s$^{-1}$, $\log_{10}\Phi$ decreases by 2.0 dex at M$_{\rm UV}=-20.25$ but only decreases by 0.3 dex at M$_{\rm UV}=-22.5$. This trend is qualitatively the same at all redshifts.

It is thus clear that a varying Ly$\alpha$ flux limit will significantly affect the UV LF as a whole, both in shape and characteristic parameters, with number densities being significantly more affected for fainter M$_{\rm UV}$. To compare UV LFs at different redshifts and interpret any evolution, it is therefore necessary to ensure we use the same luminosity ranges, otherwise a potential evolution in the UV LF of LAEs may not be intrinsic but instead could be a consequence of the different Ly$\alpha$ luminosity limits. As such, when comparing LFs, we not only compare the full samples, but also compare a homogeneous subset, defined by a single Ly$\alpha$ luminosity cut of $\log_{10} (\rm L_{Ly\alpha}/erg\,s^{-1}) \geq 43.0$, which we will apply to all redshifts. We choose this value as it excludes the lower ${\rm L_{Ly\alpha}}$ regime which can only be reached at lower redshift or by the deep NBs, and covers a luminosity regime which is probed at all redshifts, ensuring we are comparing similar samples of LAEs. While this cut will only remove a small fraction of LAEs from MBs at $z>3.5$, it will significantly reduce the number of sources at the lower redshifts, with only 10$\%$ of non-AGN LAEs at $z=2.5$ being above this Ly$\alpha$ cut.

\subsubsection{The $\log_{10} (\rm L_{Ly\alpha}/erg\,s^{-1}) \geq 43.0$ population of LAEs and limitations} \label{sec:L43}
In order to probe evolution in the same luminosity ranges, we have defined a subsample of the SC4K sample of LAEs, with $\log_{10} (\rm L_{Ly\alpha}/erg\,s^{-1}) \geq 43.0$ at all redshifts. In comparison, the characteristic L$_{\rm Ly\alpha}$ is measured to be $\log_{10} (\rm L^*_{Ly\alpha}/erg\,s^{-1})=42.93^{+0.15}_{-0.11}$ \citep{Sobral2018}, so these sources are extremely bright LAEs, rare dust-free starbursts. \cite{Amorin2017} has shown that such sources (some galaxies in that study are also selected as LAEs in the SC4K sample) are analogues of high-$z$ primeval galaxies.

Nonetheless, imposing an artificial L$_{\rm Ly\alpha}$ limit in our samples requires some caveats. As it can be seen in Fig. \ref{fig:lf_full} (right panel), even for the bright M$_{\rm UV}$ (<-21) regime, the LF only truly converges for $\log_{10} (\rm L_{Ly\alpha}/erg\,s^{-1}) < 42.7$, with a further $\sim0.1-0.2$ dex drop-off in $\Phi$ as we move to a $\sim10^{43.0}$ L$_{\rm Ly\alpha}$ cut. Introducing a L$_{\rm Ly\alpha}$ limit (corresponding to a vertical cut in Fig. \ref{fig:lya_vs_others}) introduces uncertainties in the estimation of the LFs, as with a $\log_{10} (\rm L_{Ly\alpha}/erg\,s^{-1}) \geq 43.0$ cut one can only derive the LF down to M$_{\rm UV}\sim-21$, which is comparable to M$_{\rm UV}^*$. Extremely deep surveys with e.g. MUSE can address this issue by reaching faint L$_{\rm Ly\alpha}$ even at the highest redshifts, at the cost of probing lower volumes and thus not being able to fully constrain the brightest regimes. A combined effort from IFU and NB/MB surveys \citep[see synergy/combined Ly$\alpha$ LF,][]{Sobral2018} can be the path to fully exploring the UV LFs of LAEs. In this work, while we show and discuss the best estimates computed for the samples with the $\log_{10} (\rm L_{Ly\alpha}/erg\,s^{-1}) \geq 43.0$ cut, as they provide a relative comparison of the same luminosity regimes, we focus our baseline interpretation of evolution on the full sample with no luminosity cuts.

%
%
\begin{figure*}
\begin{tabular}{ll}
\centering
  \includegraphics[width=0.487\textwidth]{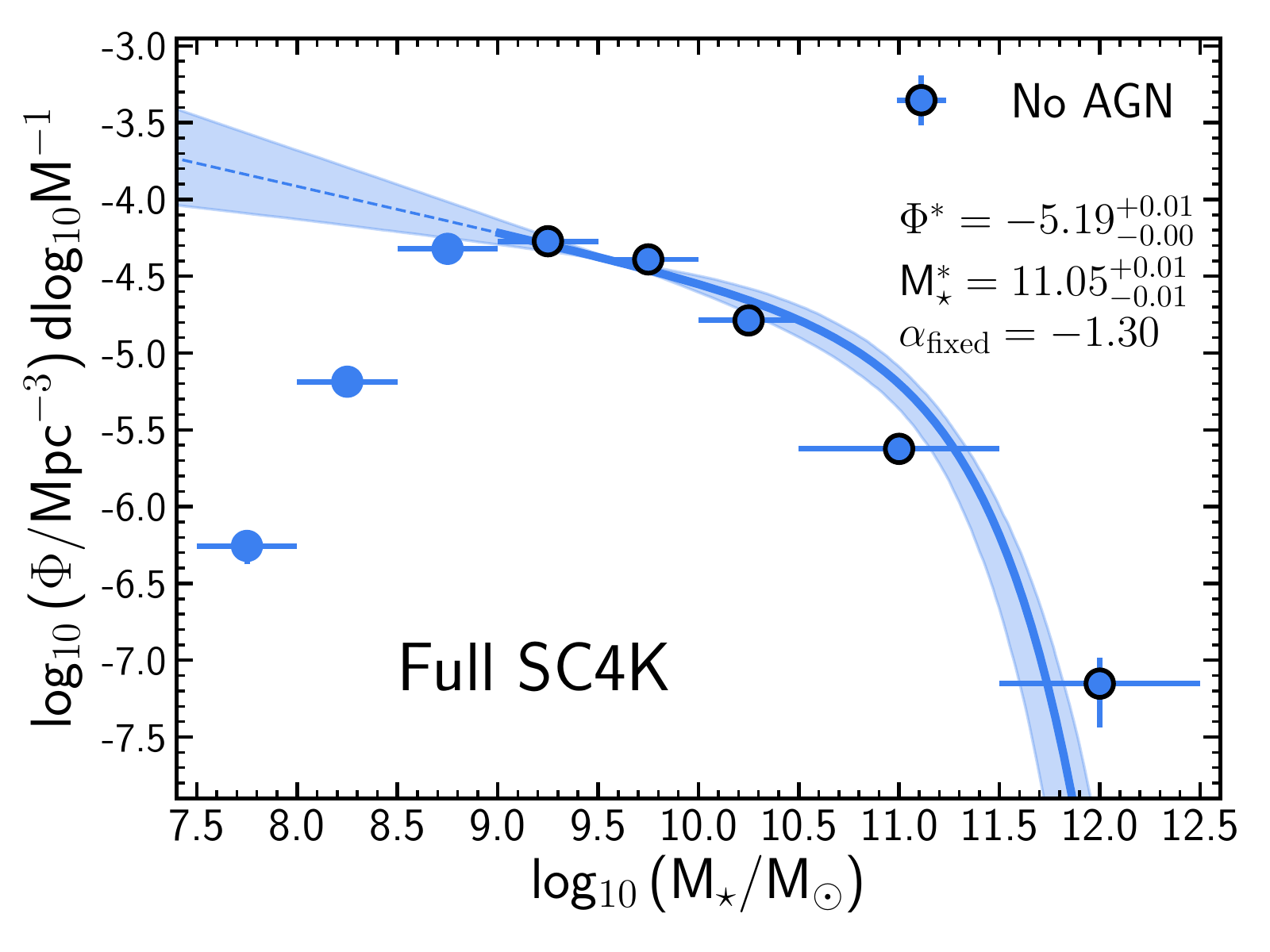}
  &
  \includegraphics[width=0.487\textwidth]{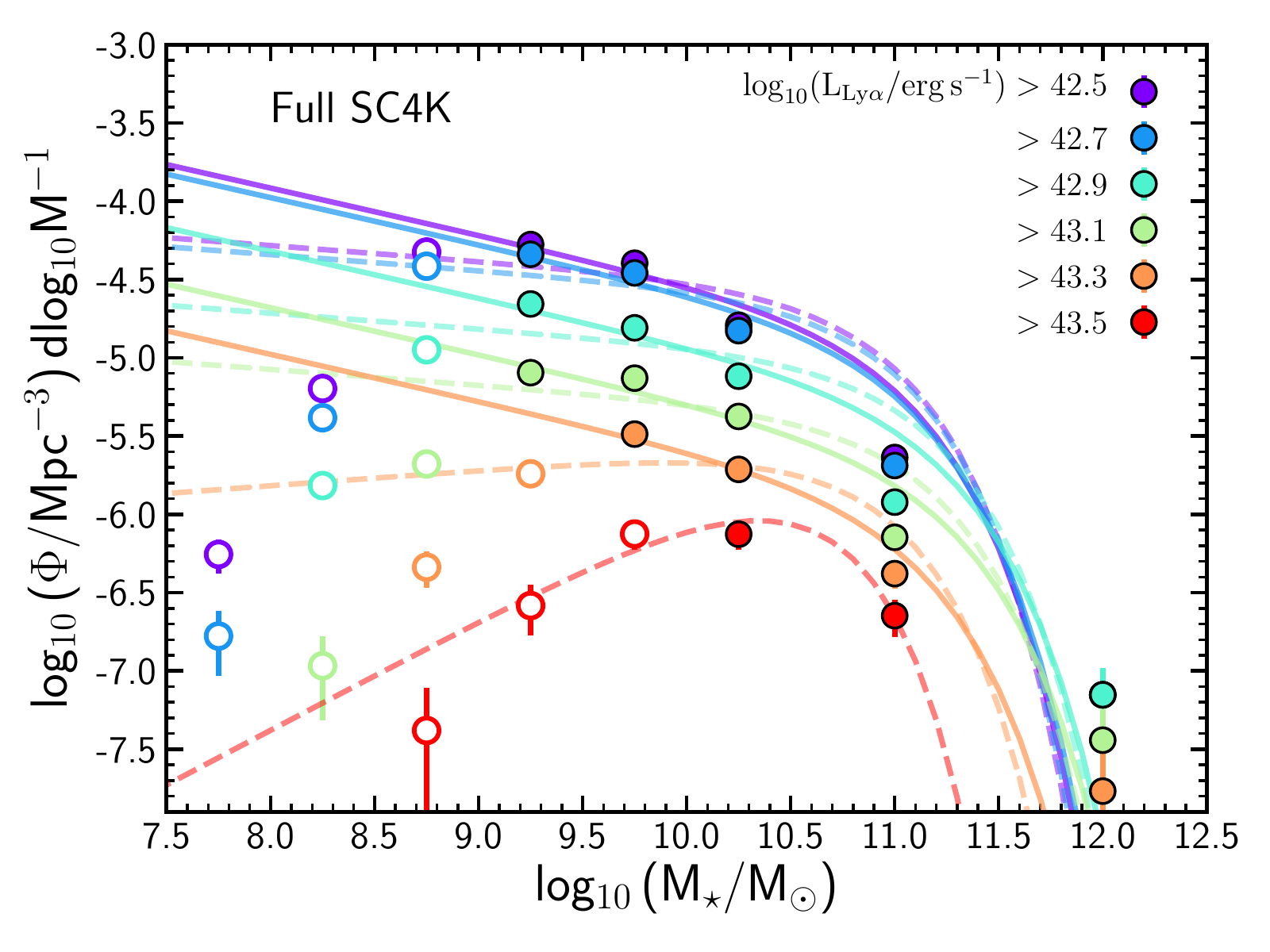}  
\end{tabular}
  \caption{{\it Left:} Stellar mass function for the full SC4K sample after removing AGN (blue points, what we use throughout this study). AGN sample is not shown here as we cannot obtain accurate mass estimations for AGNs using the stellar+dust SED-fitting we use in this study. The contours are the 16th and 84th percentiles, which we obtain by perturbing the bins within their Poissonian error bars and iterating the fitting 1000 times (see \S\ref{subsec:perturb_fits}). {\it Right:} SMF of the full SC4K sample at different L$_{\rm Ly\alpha}$ cuts. We show the best Schechter fits to the full stellar mass range as dashed lines, and to the number densities above the turn-over as filled lines (see \S\ref{subsec:smf}). The increasing L$_{\rm Ly\alpha}$ cuts reduce the number densities at all mass ranges. The decay of the number density is much more uniform across the entire mass range compared to the UV LF (Fig. \ref{fig:lf_full}, right panel), which can be explained by L$_{\rm Ly\alpha}$ and M$_\star$ having a shallower correlation with significant scatter (see Fig. \ref{fig:lya_vs_others}, right panel).}
  \label{fig:smf_full}
\end{figure*}

\subsection{Redshift evolution of the UV LF from $z\sim2$ to $z\sim6$}

We will now use our sample of LAEs, selected with 16 unique NBs and MBs in 16 well defined redshift slices, to probe the evolution of the UV LF of LAEs from $z\sim2$ to $z\sim6$. We have shown in  Fig. \ref{fig:grid_muv_lf} the UV LF for LAEs selected from each of the 16 individual NB and MB filters, together with best-fit Schechters and 1$\sigma$ countours. We provide all the Schechter parameter estimates in Table \ref{tab:schechter_params_uv_lf}. All samples are well represented by Schechter distributions. Our measurements agree well with \cite{Ouchi2008} at $z\sim3$, $z\sim4$ and $z\sim5.7$, but we report lower number densities at $z=5.8$, particularly for fainter M$_{\rm UV}$. This discrepancy can be explained by differences in Ly$\alpha$ flux limits, as the MB that we use is only sensitive to $\log_{10} (\rm L_{Ly\alpha}/erg\,s^{-1}) \geq 43.0$. We also note that our M$_{\rm UV}$ measurements are estimated from SED fitting with $30+$ bands, including the recent ultra-deep NIR data from UltraVISTA DR4, instead of directly from adjacent photometric bands.

%
%
\begin{figure*}
\begin{tabular}{ll}
  \centering
  \includegraphics[width=0.487\textwidth]{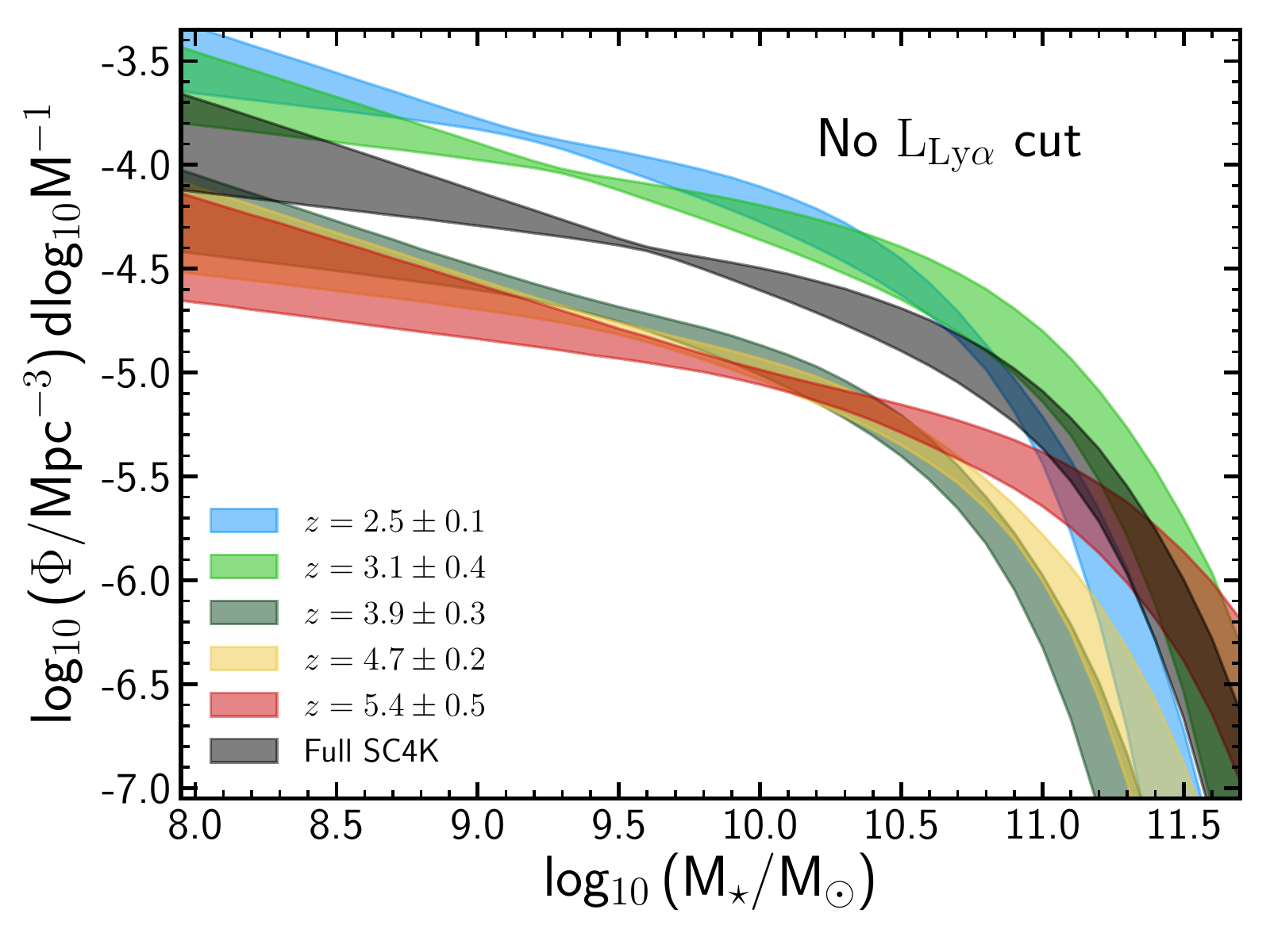}
  &
  \includegraphics[width=0.487\textwidth]{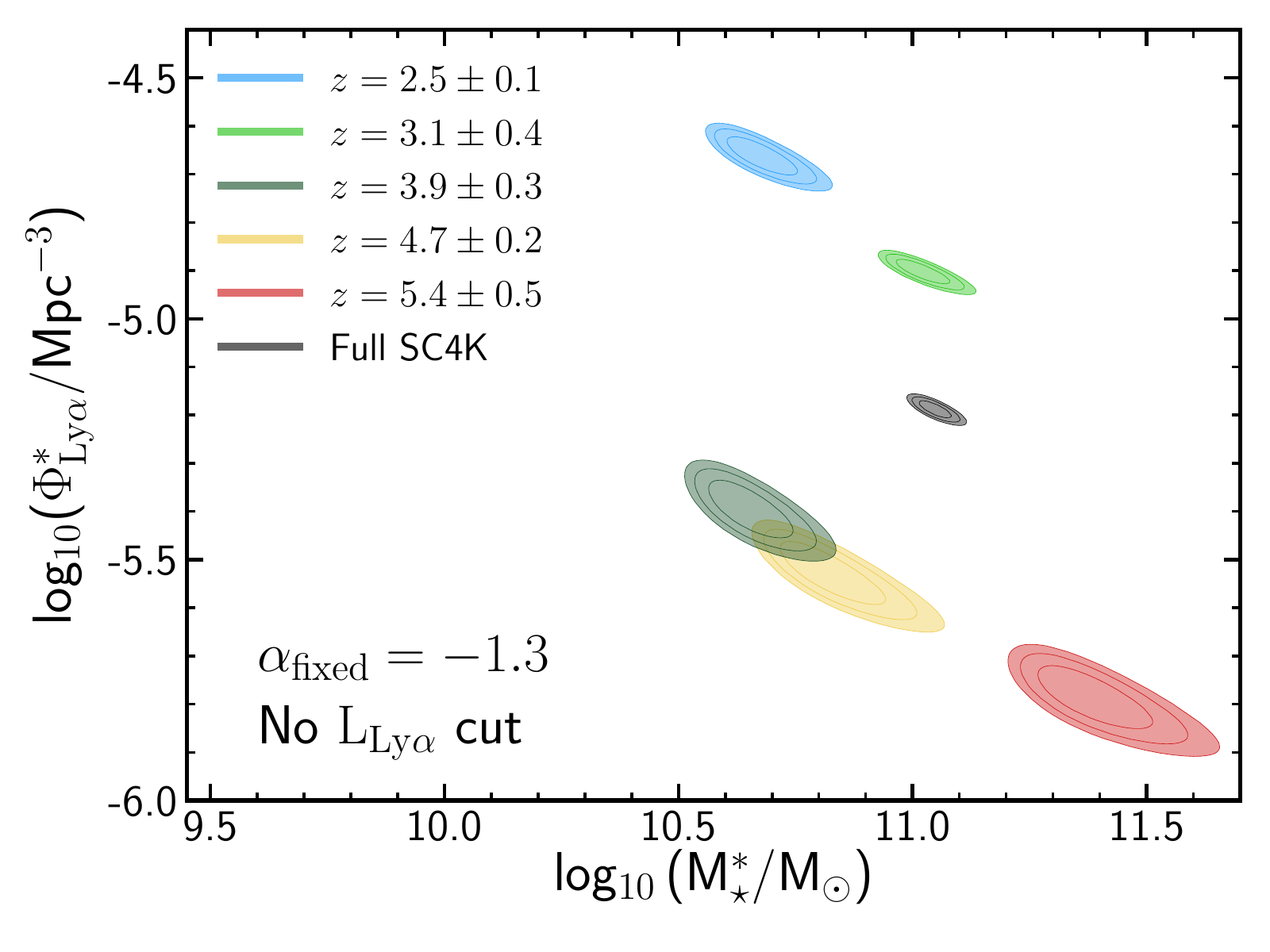}  
\end{tabular}
  \caption{{\it Left:} Evolution of the SMF with redshift, with no L$_{\rm Ly\alpha}$ cut. The shaded contours are the 16th and 84th percentiles of all iterations obtained by perturbing the luminosity bins (see \S\ref{subsec:perturb_fits}) {\it Right:} $\Phi^*-$M$_\star^*$ 1$\sigma$, 2$\sigma$ and 3$\sigma$ contours. We observe a $\log_{10}(\Phi^*/$Mpc$^{-3})$ decrease from $-4.5$ at $z=2.5$ to $-5.5$ at $z=5-6$ and $\log_{10}\,$(M$_\star^*$/M$_{\odot})$ stays constant at $\sim10.7$, although we measure small increase at $z=5.4$.}
  \label{fig:multiple_smf_no_cut}
\end{figure*}

%
%
\begin{figure*}
\begin{tabular}{ll}
  \centering
  \includegraphics[width=0.487\textwidth]{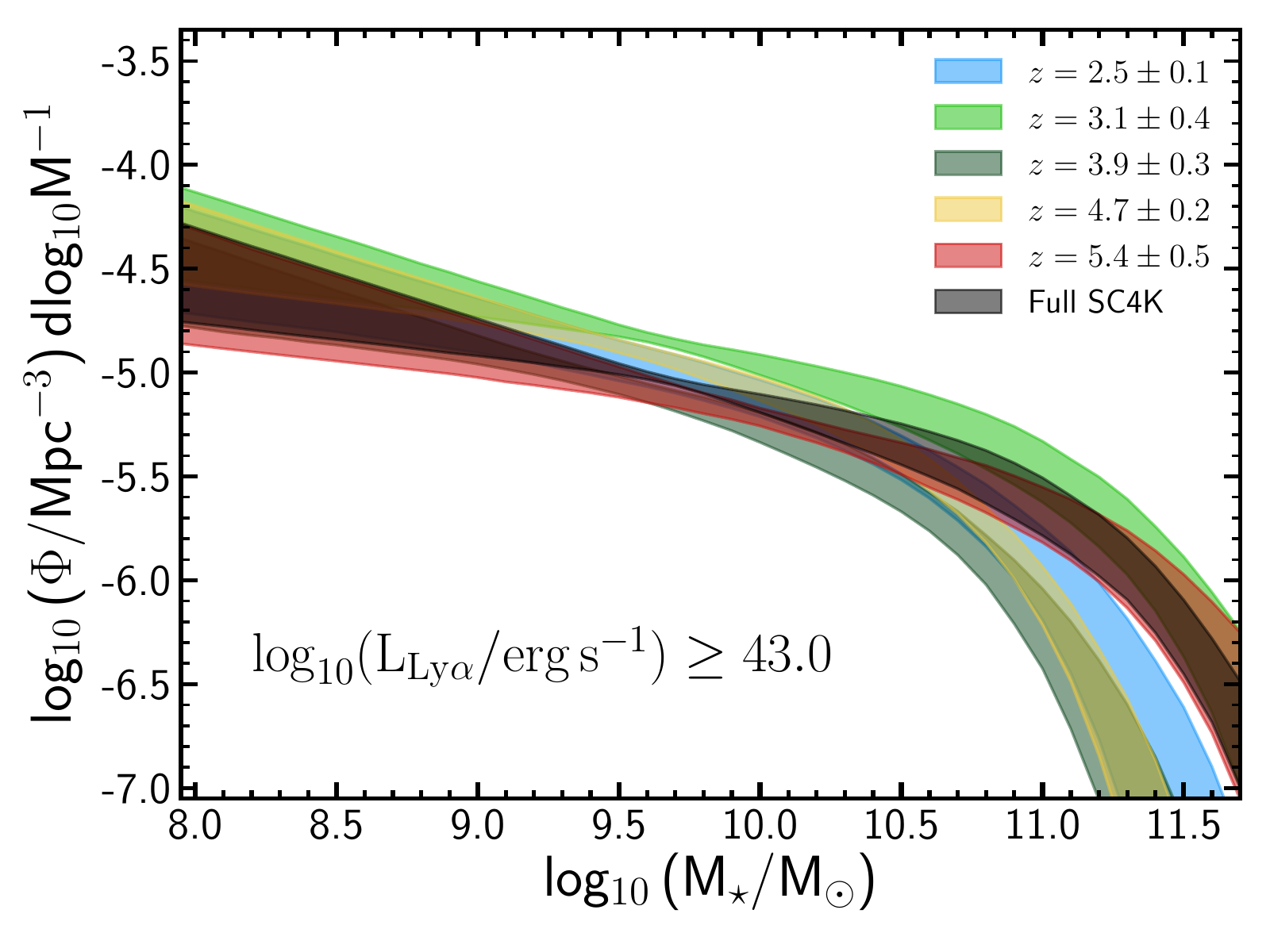}
  &
  \includegraphics[width=0.487\textwidth]{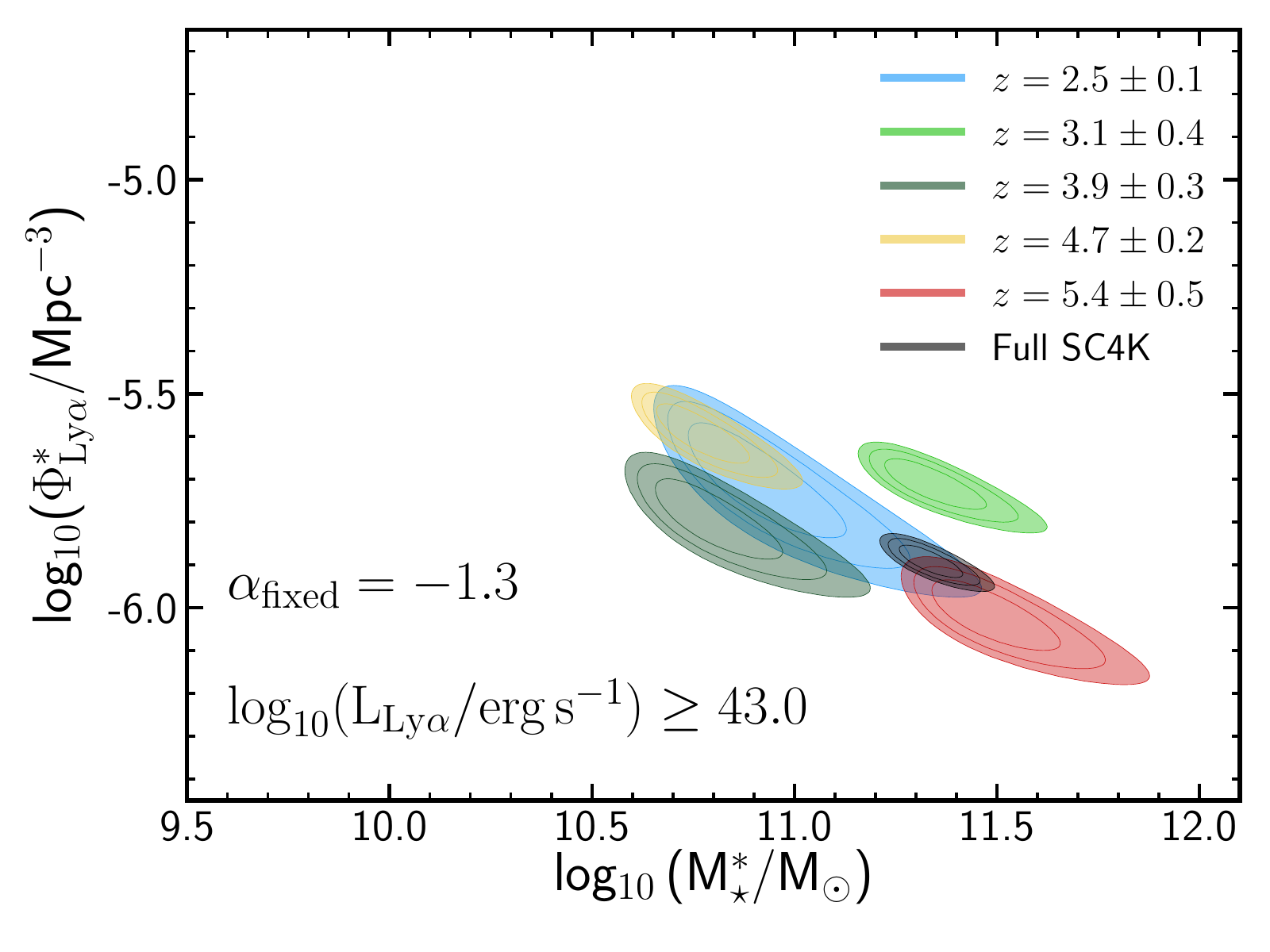}  
\end{tabular}
  \caption{{\it Left:} Evolution of the SMF with redshift, with a luminosity cut of $\log_{10} (\rm L_{Ly\alpha}/erg\,s^{-1}) \geq 43.0$. {\it Right:} $\Phi^*-$M$_\star^*$ 1$\sigma$, 2$\sigma$ and 3$\sigma$ contours. With a uniform cut for the entire sample, we do not observe clear evidence of evolution with redshift of the SMF of LAEs. We find little M$_\star^*$ and $\Phi^*$ evolution with redshift, remaining constant at $\log_{10}\,$(M$_\star^*$/M$_{\odot})\sim11$ and $\log_{10}(\Phi^*/$Mpc$^{-3})\sim5.8$.}
  \label{fig:multiple_smf}
\end{figure*}

For a statistically robust study of the evolution of UV LFs of LAEs with redshift, we group LAEs from multiple filters that probe similar redshifts to explore five different bins of redshift ($z=2.5$, $z=3.1$, $z=3.9$, $z=4.7$ and $z=5.4$; see \S\ref{subsec:binning}), as well as the full SC4K sample. The completeness corrections are applied to LAEs individually, based on their Ly$\alpha$ luminosity (see \ref{subsec:completeness}) and the volume per redshift bin is the sum of the volume of individual redshift slices included in the redshift bin (see Table \ref{tab:overview}).

We show in Fig. \ref{fig:multiple_lf_no_cut} (left panel) the UV LF at different redshifts ($z=2.5$, $z=3.1$, $z=3.9$, $z=4.7$ and $z=5.4$), without any L$_{\rm Ly\alpha}$ cut. We also show in Fig. \ref{fig:multiple_lf_no_cut} (right panel) the 1$\sigma$, 2$\sigma$ and 3$\sigma$ contours of $\Phi^*-$M$_{\rm UV}^*$. We observe a brightening (M$_{\rm UV}^*$ becomes more negative) of the UV LF with increasing redshift, from $\sim-20.5$ at $z=2.5$ to $\sim-22$ at $z=5.4$, and a $\log_{10}(\Phi^*/$Mpc$^{-3})$ decrease from $\sim-3.5$ to $\sim-4.5$ for the same redshifts. While in UV-continuum studies \citep[e.g.][]{Bouwens2015,Finkelstein2015} $\Phi^*$ of the UV LF is also measured to decrease with increasing redshift, M$_{\rm UV}^*$ is found to become fainter (increase), which is the opposite of what we measure in our sample of LAEs (before applying any luminosity cut). 

However, as previously discussed (\S\ref{sec:muv_vary}), different Ly$\alpha$ luminosity limits play a very significant role on the shape and characteristic parameters of the UV LF. We thus conduct the same analysis for a subset of our sample of LAEs, obtained by applying the luminosity cut of $\log_{10} (\rm L_{Ly\alpha}/erg\,s^{-1}) \geq 43.0$. By using a uniform cut at all redshifts (see Fig. \ref{fig:multiple_lf_cut}), we are able to probe evolution in comparable Ly$\alpha$ luminosity regimes, and reduce the effects of the Ly$\alpha$ flux limit bias, but also introducing some caveats (see discussion in \S\ref{sec:L43}). We now observe an increase of $\Phi^*$ with increasing redshift, from $\log_{10}(\Phi^*/$Mpc$^{-3})\sim-5.5$ at $z=2.5$ to $-4.5$ at $z\sim3-6$, which contrasts the decrease observed in UV-continuum selected samples. We do not observe trends in M$_{\rm UV}^*$ evolution, which also contrasts the increase in M$_{\rm UV}^*$ observed in UV-continuum selected samples. There is no evolution of $\log_{10}(\Phi^*/$Mpc$^{-3})$ between $z\sim3$ and $z\sim6$, but we observe a brightening between $z\sim3$ and $z\sim5-6$, which is the same trend reported by \cite{Ouchi2008}.

\subsection{The global SMF of LAEs at $z\sim2-6$}

Following the same methodology that we use for the UV LF, we now analyse the global SMF of $\sim4000$ LAEs at $z\sim2-6$. The study of the SMF of such a large sample of LAEs over such a wide volume is unprecedented at these redshift ranges. We have a robust sample of LAEs at $10^{9.0}-10^{12.5}$\,M$_\odot$, with individual measurements down to $\sim10^{7.5}$\,M$_\odot$. Studies that have estimated stellar masses of $z>2$ galaxies, typically only probe $>10^{10}$\,M$_\odot$ galaxies \citep[e.g.][]{Schreiber2015} but with our population of LAEs, we are capable of reaching galaxies with very low stellar masses, while still having detections of very massive systems ($>10^{11}$\,M$_\odot$). 

We show in Fig. \ref{fig:smf_full} (left panel) the SMF of the full SC4K sample of $z\sim2-6$ LAEs after removing AGN (which is what we use throughout this paper, see \S\ref{sec:sample_AGNs}). Unlike the UV LF, we do not explore how AGNs influence the SMF since we are not able to accurately estimate the stellar mass of AGNs with our stellar$+$dust SED-fitting code which does not use AGN models. We show the Schechter fit to the SMF and the 1$\sigma$ contour which we estimate by perturbing the stellar mass bins within their Poissonian errors and fitting 1000 realisations of the perturbed bins (see \S\ref{subsec:perturb_fits}). The SMF resembles a Schechter distribution, but with an excess in number densities at $10^{12}$\,M$_\odot$.

%
%
\begin{figure*}
  \centering
  \includegraphics[width=\textwidth]{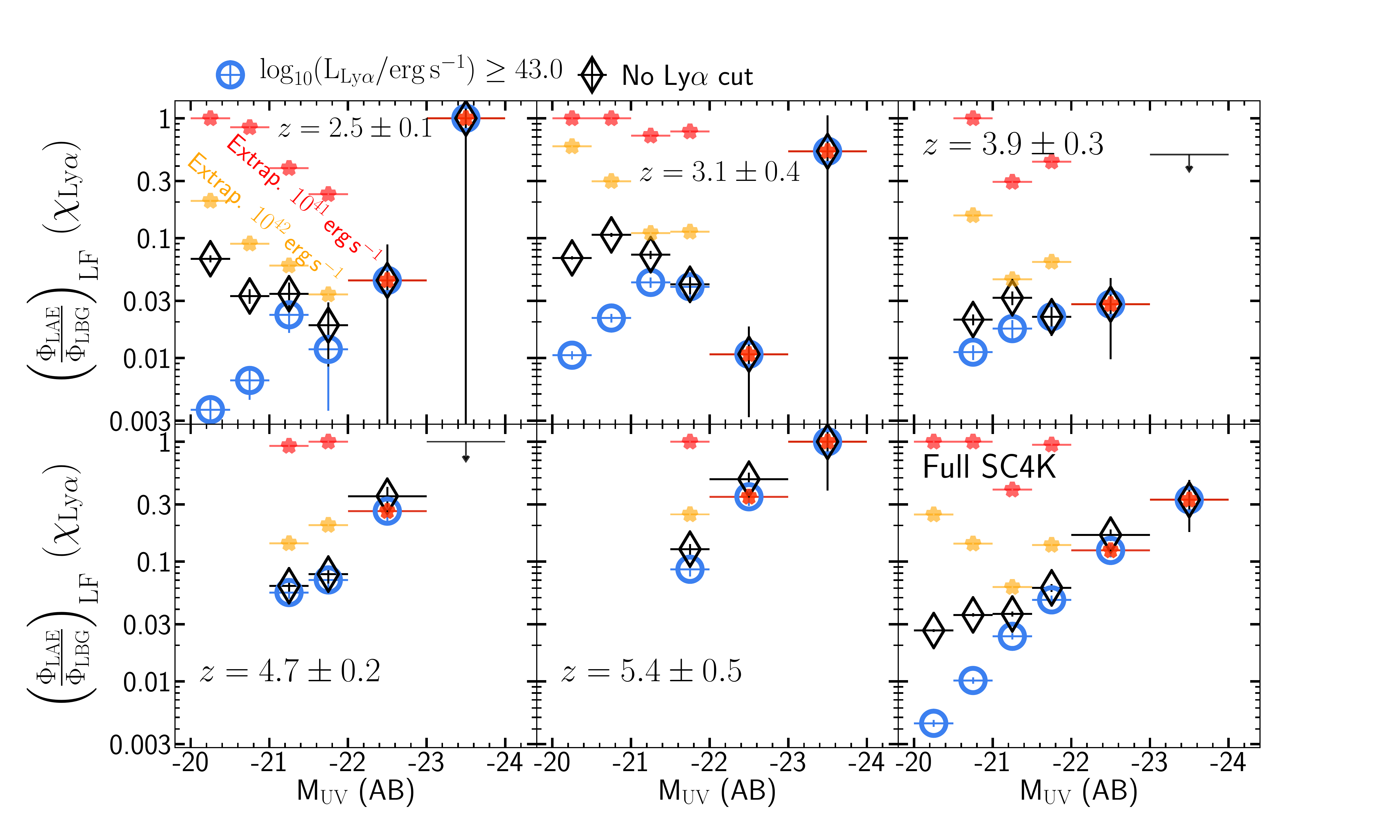}
  \caption{$\Phi_{\rm LAE}/\Phi_{\rm LBG}$ ratio (interpreted as $\chi_{\rm Ly\alpha}$) dependence on M$_{\rm UV}$ for different redshifts. $\Phi_{\rm LAE}/\Phi_{\rm LBG}$ measurements are shown when applying a uniform $\log_{10} (\rm L_{Ly\alpha}/erg\,s^{-1}) \geq 43.0$ cut (blue circles), and when applying no cut (black diamonds). The ratio is computed from a compilation of UV LFs from UV-selected galaxies at $z=2.3$, $z=3.05$ \citep{Reddy2009}, $z=4$ , $z=5$ and $z=6$ \citep{Ono2018}. We show simple extrapolations to $\log_{10} (\rm L_{Ly\alpha}/erg\,s^{-1}) \geq 42.0$ (orange stars) and $\log_{10} (\rm L_{Ly\alpha}/erg\,s^{-1}) \geq 41.0$ (red stars), computed from $z=2.5$ and applied to all redshift intervals. For better visualisation, the ratio is collapsed to $\Phi_{\rm LAE}/\Phi_{\rm LBG}=1$ when it surpasses that value.}
  \label{fig:lya_fraction_muv}
\end{figure*}

\subsection{SMF with varying {\rm L$_{\rm Ly\alpha}$} cuts} \label{sec:smf_vary}

Here, we explore how different Ly$\alpha$ luminosity limits affect the SMF. For the UV LF of LAEs, we have observed that an increasing L$_{\rm Ly\alpha}$ cut significantly affects the shape and characteristic parameters of the distribution, with a more significant effect on the number density of fainter UV luminosities, which are typically linked with lower Ly$\alpha$ luminosities. Such a trend is not necessarily expected for the SMF, as the relation between M$_\star$ and L$_{\rm Ly\alpha}$ is very shallow, if even present (see Fig. \ref{fig:lya_vs_others}, right panel).

We show in Fig. \ref{fig:smf_full} (right panel) the SMF of the full SC4K sample with varying ${\rm L_{Ly\alpha}}$ cuts, from $10^{42.5}$ to $10^{43.5}$\,erg\,s$^{-1}$. As the stellar mass and L$_{\rm Ly\alpha}$ have a shallow relation, an increasing ${\rm L_{Ly\alpha}}$ produces a much more uniform decay of the number densities over the entire stellar mass range. Between $10^{42.5}$ and $10^{43.3}$\,erg\,s$^{-1}$, $\log_{10}(\Phi^*/$Mpc$^{-3})$ decreases by 1.6 dex at $\log_{10}\,$(M$_\star$/M$_{\odot})=9.25$ and by 1.0 dex at $\log_{10}\,$(M$_\star$/M$_{\odot})=11.0$, which is much more modest than the large difference observed for the UV LF.

As such, when comparing SMFs at different redshifts, we will not only look at the full samples, but we will also make use of a luminosity cut $\log_{10} (\rm L_{Ly\alpha}/erg\,s^{-1}) \geq 43.0$, for the same reasons that we do for the UV LF (see the discussion in \S\ref{sec:muv_vary}, including the limitations associated with fitting a LF to our sample after applying a luminosity cut). This produces a luminosity range which all filters can target and is consistent with our approach to compare UV LFs.

%
%
\begin{figure*}
  \centering
  \includegraphics[width=0.8\textwidth]{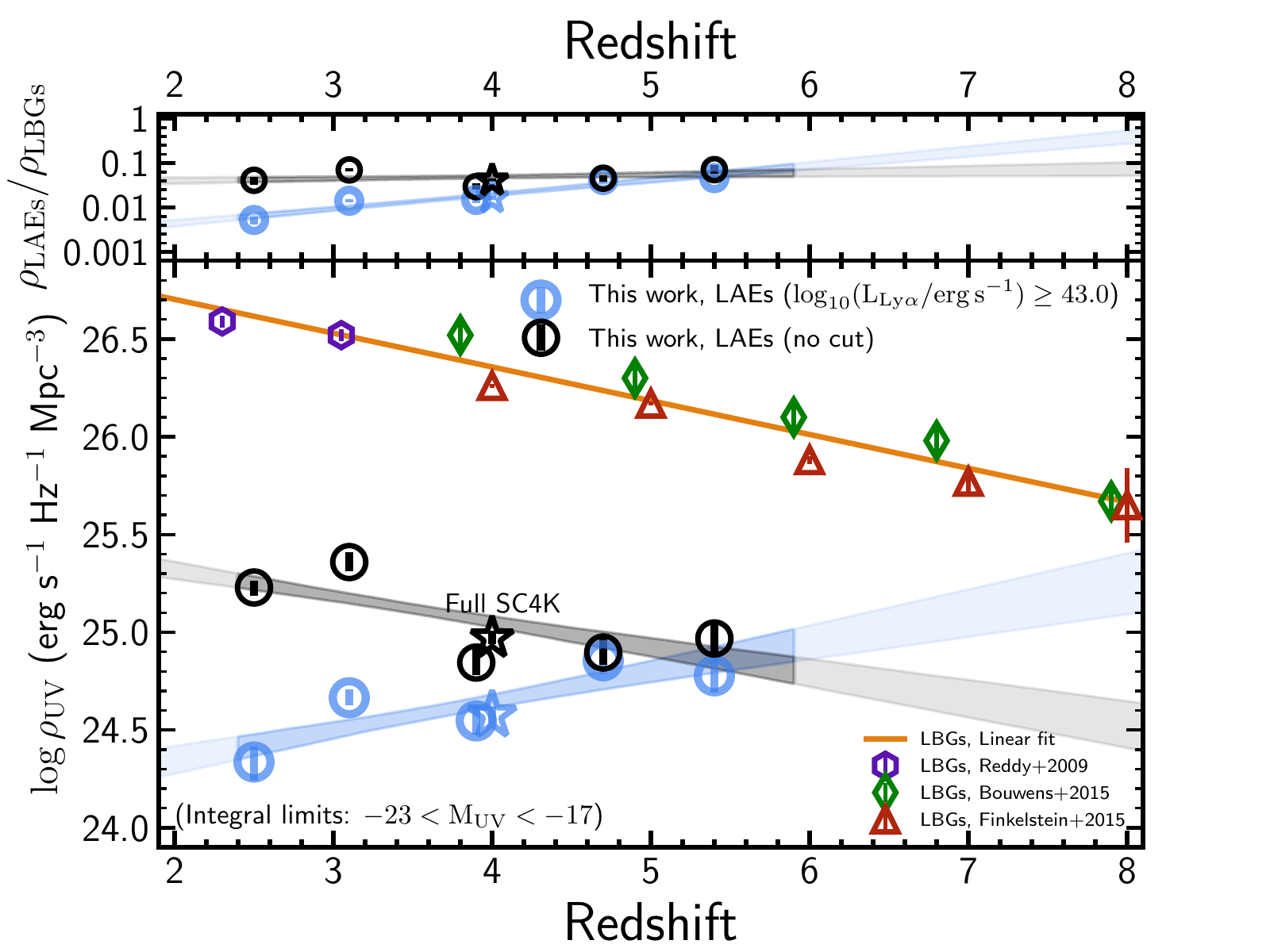}
  \caption{Evolution of the UV luminosity ($\rho_{\rm UV}$) with redshift. We show the $\rho_{\rm UV}$ measurements for our sample of LAEs (black circles), but also with a consistent $\log_{10} (\rm L_{Ly\alpha}/erg\,s^{-1}) \geq 43.0$ cut (blue circles, see discussion on limitations of fitting the LF of this subsample in \S\ref{sec:L43}). $\rho_{\rm UV}$ of the full SC4K sample is shown as stars, using the same colour scheme. The shaded contours are the 16th and 84th percentiles of the fits, obtained by perturbing the M$_{\rm UV}$ bins at each redshift (see \S\ref{subsec:perturb_fits}). We find no evidence for $\rho_{\rm M}$ evolution with redshift when applying a consistent ${\rm L_{Ly\alpha}}$ cut. We compare our results with measurements from the literature, from continuum-selected LBG populations: $z=2.3$, $z=3.05$ \citep{Reddy2009}, $z=3.8$ , $z=4.9$, $z=5.9$, $z=6.8$, $z=7.9$ \citep{Bouwens2015}, $z=4$, $z=5$, $z=6$, $z=7$, $z=8$ \citep{Finkelstein2015}.}
  \label{fig:SMD}
\end{figure*}

%
%
\begin{figure*}
  \centering
  \includegraphics[width=0.8\textwidth]{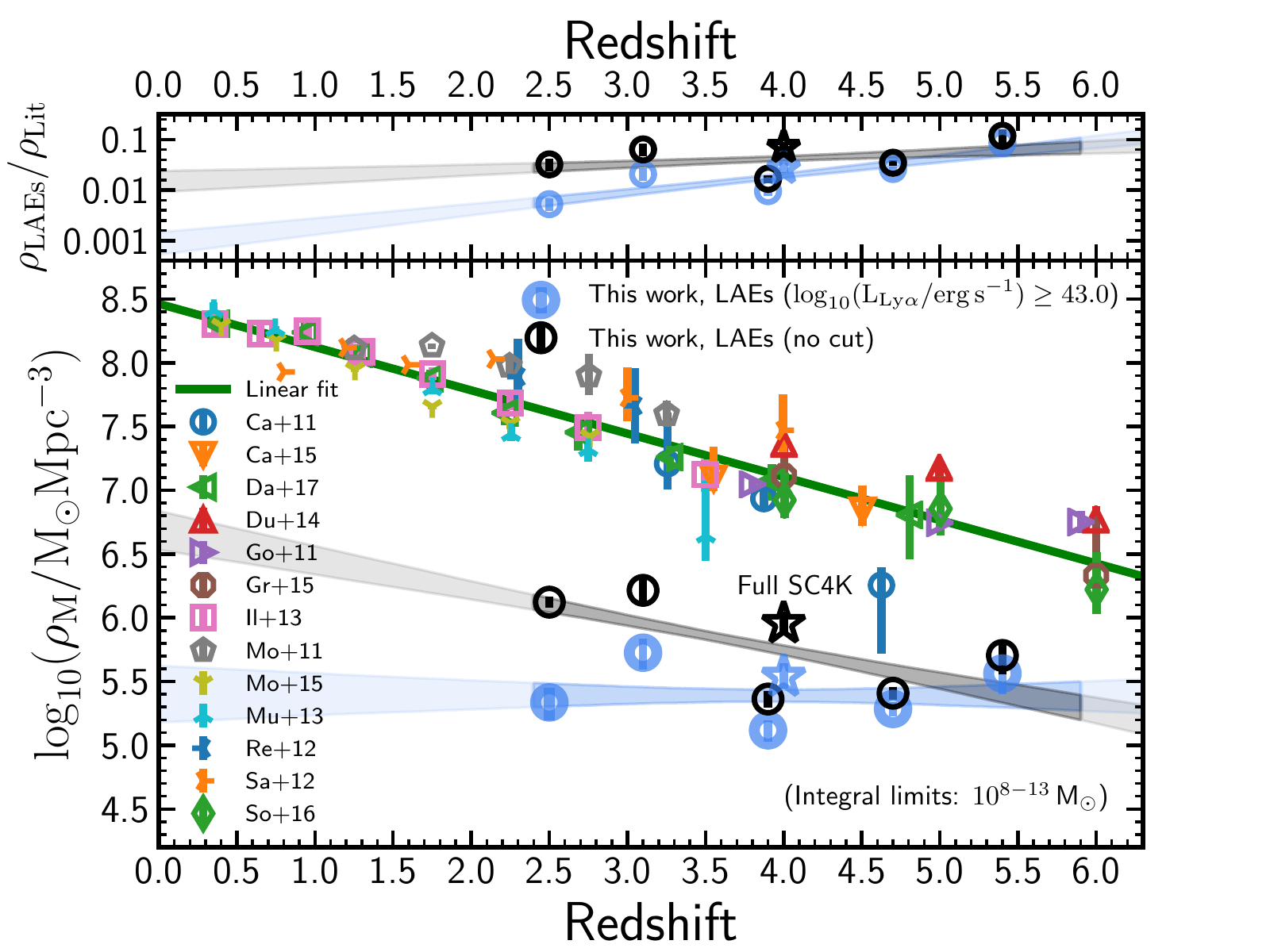}
  \caption{Evolution of the stellar mass density ($\rho_{\rm M}$) with redshift. We show the $\rho_{\rm M}$ measurements for our sample of LAEs (black circles), but also with a consistent $\log_{10} (\rm L_{Ly\alpha}/erg\,s^{-1}) \geq 43.0$ cut (blue circles, see discussion on limitations of fitting LFs of this subsample in \S\ref{sec:L43}). $\rho_{\rm M}$ of the full SC4K sample is shown as stars, using the same colour scheme. The shaded contours are the 16th and 84th percentiles of the fits, obtained by perturbing the stellar mass bins at each redshift (see \S\ref{subsec:perturb_fits}). We find no evidence for $\rho_{\rm M}$ evolution with redshift when applying a consistent ${\rm L_{Ly\alpha}}$ cut. We compare our results with measurements from the literature, from continuum-selected populations: \citealt{Davidzon2017} (Da+17), \citealt{Caputi2011} (Ca+11), \citealt{Caputi2015} (Ca+15), \citealt{Duncan2014} (Du+14), \citealt{Gonzalez2011} (Go+11), \citealt{Grazian2015} (Gr+15), \citealt{Ilbert2013} (Il+13), \citealt{Mortlock2011} (Mo+11), \citealt{Mortlock2015} (Mo+15), \citealt{Muzzin2013} (Mu+13), \citealt{Reddy2012} (Re+12), \citealt{Santini2012} (Sa+12), \citealt{Song2016} (So+16), and \citealt{Tomczak2014} (To+14). All $\rho_{\rm M}$ were converted to Chabrier, when another IMF was used. We show the best fit to this compilation as a green line. The ratio between the $\rho_{\rm M}$ from the literature and $\rho_{\rm M}$ from $\log_{10} (\rm L_{Ly\alpha}/erg\,s^{-1}) \geq 43.0$ LAEs (top panel) decreases from $\sim300$ at $z\sim2.5$ to $\sim30$ at $z\sim5-6$, suggesting an increasing overlap between populations with increasing redshift.}
  \label{fig:SMD}
\end{figure*}

\subsection{Redshift evolution of the SMF of LAEs from $z\sim2$ to $z\sim6$}

We probe the evolution of the SMF with redshift, using $\sim4000$ LAEs selected in 16 well defined redshift slices from $z\sim2$ to $z\sim6$. We showed the SMF of LAEs selected from individual filters in Fig. \ref{fig:grid_smf}, together with 1$\sigma$ Schechter contours. All redshift slices resemble a Schechter distribution and we provide the best-fit parameters in Table \ref{tab:schechter_params_smf}.

In order to obtain statistically robust comparisons of the evolution of the SMF of LAEs with redshift, we follow the same grouping scheme that we use for the UV LFs. We define five redshift intervals ($z=2.5$, $z=3.1$, $z=3.9$, $z=4.7$ and $z=5.4$; see \S\ref{subsec:binning}) and also use the global SMF of the full $z\sim2-6$ sample. The completeness corrections are applied to LAEs individually, based on their Ly$\alpha$ luminosity (see \ref{subsec:completeness}) and the volume per redshift bin is the sum of the volume of individual redshift slices included in the redshift bin (see Table \ref{tab:overview}). We show in Fig. \ref{fig:multiple_smf_no_cut} (left panel) the SMF at different redshifts ($z=2.5$, $z=3.1$, $z=3.9$, $z=4.7$ and $z=5.4$), without any L$_{\rm Ly\alpha}$ cut. We also show in Fig. \ref{fig:multiple_smf_no_cut} (right panel) the 1$\sigma$, 2$\sigma$ and 3$\sigma$ contours of $\Phi^*-$M$_\star^*$. We observe a clear evolution of the SMF with redshift (before applying any Ly$\alpha$ luminosity restriction), with the low mass end shifting down by 1 dex from $z=2.5$ to $z=5.4$. This is reflected as a gradual $\log_{10}(\Phi^*/$Mpc$^{-3})$ decrease with redshift from $-4.6$ at $z=2.5$ to $-5.8$ at $z=5.4$. We measure an M$_\star^*$ increase from $\log_{10}(\rm M_\star^*/M_\odot)=10.6$ to 11.5 at the same redshift ranges. The shift down to lower $\Phi^*$ with increasing redshift is also observed in the SMF of more typical galaxies \citep[e.g.][]{Muzzin2013}, which suggests the observed trends are qualitatively the same, however, an analysis using the same luminosity regime is still required.

As previously discussed in \S\ref{sec:smf_vary}, different Ly$\alpha$ luminosity limits play a very significant role on the shape and characteristic parameters of the SMF. We thus conduct the same analysis for a subset of our sample of LAEs, obtained by applying the luminosity cut of $\log_{10} (\rm L_{Ly\alpha}/erg\,s^{-1}) \geq 43.0$. By using a uniform cut at all redshifts (see Fig. \ref{fig:multiple_smf}), we are able to probe evolution in comparable Ly$\alpha$ luminosity regimes, and reduce the effects of the Ly$\alpha$ flux limit bias, but also introducing some caveats (see discussion in \S\ref{sec:L43} for the UV LF, as the limitations raised also apply to the SMF). While there is a clear evolution in the observed Schechter fits of the full samples, we find no evidence of such evolution when comparing samples of LAEs within the same Ly$\alpha$ regime. We find little M$_\star^*$ and $\Phi^*$ evolution with redshift, remaining constant at $\log_{10}\,$(M$_\star^*$/M$_{\odot})\sim11$ and $\log_{10}(\Phi^*/$Mpc$^{-3})\sim-5.8$. The evolution that we find when looking at the same luminosity regimes is thus not qualitatively the same that is observed in more typical galaxies. Analysis of the evolution of the stellar mass density, will provide more insight into this.

\subsection{Evolution of the Ly$\alpha$ fraction}  \label{sec:lya_frac_evo}

We attempt to infer the Ly$\alpha$ fraction ($\chi_{\rm Ly\alpha}$) dependence on redshift and M$_{\rm UV}$. We compute the ratio between the observed UV number densities in our sample of LAEs and the UV number densities of LBGs from the literature: $\Phi_{\rm LAE}/\Phi_{\rm LBG}$, which can be interpreted as the fraction of LBGs that are LAEs (above some Ly$\alpha$ detection limit), or $\chi_{\rm Ly\alpha}$. To compute this fraction, we use a UV LF compilation consisting of: $z=2.3$, $z=3.05$ \citep{Reddy2009}, $z=4$, $z=5$ and $z=6$ \citep{Ono2018} (which we use for the redshifts $z=2.5$, $z=3.1$, $z=3.9$, $z=4.7$ and $z=5.4$, respectively). For the full SC4K sample (median $z=4.1$) we use the $z=4$ literature measurements from \citep{Ono2018}, which being a very wide area LBG survey, provides a fair comparison with our wide area LAE survey. To prevent any biases from fitting, the ratio is computed directly from the luminosity bins in this study and the literature, with the latter being interpolated to the M$_{\rm UV}$ values used in this study.

As clearly seen for the full sample in Fig. \ref{fig:lf_full} (left panel), the number density of faint M$_{\rm UV}$ LBGs is multiple times higher than the number density of faint M$_{\rm UV}$ LAEs. The number densities of M$_{\rm UV}=-20$ LAEs is $\sim1.5$ dex lower than LBGs, but they converge to the same number densities for M$_{\rm UV}$ brighter than -23. We show the ratio of the two number densities in Fig. \ref{fig:lya_fraction_muv} for five redshift intervals (and the full SC4K sample), before and after applying a L$_{\rm Ly\alpha}$ cut. The $z=2.5$ panel shows that as we probe fainter Ly$\alpha$ luminosities, we get closer to unity in the $\Phi_{\rm LAE}/\Phi_{\rm LBG}$ fraction, and that the effect of the Ly$\alpha$ cut depends on M$_{\rm UV}$, as shown in \S\ref{sec:muv_vary}. For very bright M$_{\rm UV}$ ($<-23$), we are always able to retrieve most galaxies, as M$_{\rm UV}$ bright are typically also Ly$\alpha$ bright (see Fig. \ref{fig:lya_vs_others}, left panel) and the $\Phi_{\rm LAE}/\Phi_{\rm LBG}$ will always be close to unity. This holds true for all redshifts, with the ratio always tending to unity at the brightest UV luminosities. When comparing $\Phi_{\rm LAE}/\Phi_{\rm LBG}$ at different redshifts, for the comparable $\log_{10} (\rm L_{Ly\alpha}/erg\,s^{-1}) \geq 43.0$ subsample, we observe that $\Phi_{\rm LAE}/\Phi_{\rm LBG}$ is typically higher at $z>4$ than for the lower redshift samples. This may imply that LAEs become a bigger subset of LBGs with increasing redshift \citep[same trend found in e.g.][]{Arrabal2020}, but we explore this further by measuring the UV luminosity density.

We make a direct extrapolation of the measurements of $\log_{10} (\rm L_{Ly\alpha}/erg\,s^{-1}) \geq 43.0$ and $\geq 42.5$ $z=2.5$ LAEs to lower L$_{\rm Ly\alpha}$ cuts by scaling the increment in $\Phi_{\rm LAE}$. The extrapolated values for $\log_{10} (\rm L_{Ly\alpha}/erg\,s^{-1}) \geq 42.0$ and $\geq 41.0$ are shown in Fig. \ref{fig:lya_fraction_muv}. We find that for M$_{\rm UV}=-20$ at $z=2.5$, we would approach unit if we could reach $\log_{10} (\rm L_{Ly\alpha}/erg\,s^{-1})=41.0$. We make the simple assumption that the extrapolation we predict for $z=2.5$ is valid for all redshifts, as the higher flux limits of the other redshifts are not capable of reaching $\log_{10} (\rm L_{Ly\alpha}/erg\,s^{-1}) \geq 42.5$ and thus do not allow a direct extrapolation. We find that for $z\gtrsim3$ the ratio approaches unit even for M$_{\rm UV}=-21$ to $-22$. We note that for $z>4$ and for the full SC4K sample, the extrapolation at M$_{\rm UV}=-22.5$ can be below the measurement without applying any Ly$\alpha$ cut, which is a consequence of applying the $z=2.5$ extrapolation estimation, which has a null increment for that M$_{\rm UV}$ value.

\subsection{Redshift evolution of the UV luminosity density of $z\sim2-6$ LAEs}  \label{sec:smd_evo}

We measure the UV luminosity density ($\rho_{\rm UV}$) at the aforementioned redshift intervals in our sample of LAEs and explore its evolution. We detail how the integration is conducted in \S\ref{subsec:methods_densities}, with $\alpha$ being fixed to -1.5 but perturbed within $\pm$0.2 dex. We provide our $\rho_{\rm UV}$ measurements in Table \ref{tab:rho}.

%
%
\begin{table}
\setlength{\tabcolsep}{4pt}
\begin{center}
\caption{$\rho_{\rm UV}$ and $\rho_{\rm M}$ computed for the full samples of LAEs and for a subsample with a luminosity cut of $\log_{10} (\rm L_{Ly\alpha}/erg\,s^{-1}) \geq 43.0$. We detail the integration process in \S\ref{subsec:methods_densities}. $\rho'_{UV}=\rm\rho_{UV}/erg\,s^{-1}\,Hz^{-1}\,Mpc^{-3}$; $\rho'_{M}=\rm\rho_{M}/M_\odot\,Mpc^{-3}$.} \label{tab:rho}
\begin{tabular}{c | cccc}
\hline
\multicolumn{1}{c|}{Redshift} &
\multicolumn{1}{c|}{$\log_{10}\rm\rho'_{UV}$} &
\multicolumn{1}{c|}{$\log_{10}\rm\rho'_{UV}$} &
\multicolumn{1}{c|}{$\log_{10}\rm\rho'_{M}$} &
\multicolumn{1}{c|}{$\log_{10}\rm\rho'_{M}$} \\
&& $(\log\rm L\geq43)$& &$(\rm\log L\geq43)$\\
\hline
$2.5\pm0.1$ & $25.23^{+0.03}_{-0.04}$ & $24.34^{+0.08}_{-0.10}$ & $6.12^{+0.05}_{-0.04}$ & $5.34^{+0.11}_{-0.14}$\\
$3.1\pm0.4$ & $25.36^{+0.05}_{-0.05}$ & $24.66^{+0.04}_{-0.04}$ & $6.22^{+0.13}_{-0.10}$ & $5.73^{+0.11}_{-0.13}$\\
$3.9\pm0.3$ & $24.84^{+0.07}_{-0.07}$ & $24.55^{+0.08}_{-0.07}$ & $5.36^{+0.06}_{-0.07}$ & $5.12^{+0.08}_{-0.09}$\\
$4.7\pm0.2$ & $24.90^{+0.07}_{-0.06}$ & $24.85^{+0.07}_{-0.07}$ & $5.41^{+0.05}_{-0.05}$ & $5.28^{+0.05}_{-0.06}$\\
$5.4\pm0.5$ & $24.97^{+0.09}_{-0.08}$ & $24.78^{+0.10}_{-0.09}$ & $5.70^{+0.11}_{-0.15}$ & $5.56^{+0.13}_{-0.16}$\\
\hline
Full SC4K & $24.97^{+0.04}_{-0.03}$ & $24.58^{+0.04}_{-0.04}$ & $5.96^{+0.09}_{-0.08}$ & $5.54^{+0.10}_{-0.09}$\\
\hline
\end{tabular}
\end{center}
\end{table}

When applying no luminosity restriction, we measure that $\log_{10}(\rm\rho_{UV}/erg\,s^{-1}\,Hz^{-1}\,Mpc^{-3})$ is anti-correlated with redshift, with a moderate decline from 25.2 at $z=2.5$ to 25.0 at $z\sim5-6$. When applying the luminosity cut of $\log_{10} (\rm L_{Ly\alpha}/erg\,s^{-1}) \geq 43.0$,  $\log_{10}(\rm\rho_{UV}/erg\,s^{-1}\,Hz^{-1}\,Mpc^{-3})$ of LAEs changes from 24.3 to 25.0. In comparison, $\log_{10}(\rm\rho_{UV}/erg\,s^{-1}\,Hz^{-1}\,Mpc^{-3})$ of LBGs is always higher and decreases with redshift, from 26.5 at $z=2.5$ to 26.0 at $z=6$. We extrapolate the ratio between the luminosity densities of $\log_{10} (\rm L_{Ly\alpha}/erg\,s^{-1}) \geq 43.0$ LAEs and LBGs and determine it tends to unity at $z=9$. Overall, our measurements of $\rho_{\rm UV}$ suggest that at $z\sim2$ LAEs constitute a much smaller subset of LBGs and that with increasing redshift, both populations slowly approach the same values of 
$\rho_{\rm UV}$. This is qualitatively similar to the trends found by \cite{Sobral2018} by integrating Ly$\alpha$ LFs.

\subsection{Redshift evolution of the stellar mass density of $z\sim2-6$ LAEs}  \label{sec:smd_evo}

Using our best-derived fits (Table \ref{tab:schechter_params_smf}), we estimate the stellar mass density ($\rho_{\rm M}$) of our LAEs at different redshifts, by integrating the SMFs in the range. We obtain $\rho_{\rm M}$ using the procedure described in \S\ref{subsec:methods_densities}.  We provide our $\rho_{\rm M}$ measurements in Table \ref{tab:rho}. In Fig. \ref{fig:SMD}, we show our $\rho_{\rm M}$ measurements and compare them with measurements from the literature. The observed $\rho_{\rm M}$ (without applying any luminosity cuts) changes from $\log_{10}(\rm\rho_{M}/M_\odot\,Mpc^{-3})\sim$ 6.1 at $z\sim2.5$ to $\sim5.5$ at $z\sim5-6$. By applying the consistent $\log_{10} (\rm L_{Ly\alpha}/erg\,s^{-1}) \geq 43.0$ cut, the estimated $\rho_{\rm M}$ of our LAE sample remains roughly constant with redshift at $\log_{10}(\rm\rho_{M}/M_\odot\,Mpc^{-3})\sim5.5$. 

We compare our results with measurements from the literature, from continuum-selected populations: \cite{Davidzon2017}, \cite{Caputi2011}, \cite{Caputi2015}, \citealt{Duncan2014}, \cite{Gonzalez2011}, \cite{Grazian2015}, \cite{Ilbert2013}, \cite{Mortlock2011}, \cite{Mortlock2015}, \cite{Muzzin2013}, \cite{Reddy2012}, \cite{Santini2012}, \cite{Song2016}, and \cite{Tomczak2014}. The $\rho_{\rm M}$ measurements of typical populations of galaxies from the literature indicate a decrease from $\log_{10}(\rm\rho_{M}/M_\odot\,Mpc^{-3})\sim7.5$ at $z\sim2.5$ to $\sim6.5$ at $z\sim5-6$. This implies that galaxies selected as LAEs always have low stellar mass densities, and as we move to higher redshifts, their properties become similar to the ones derived from more typical populations of galaxies, suggesting that with an increasing redshift more galaxies become LAE-like. The ratio between the stellar mass densities for the $\log_{10} (\rm L_{Ly\alpha}/erg\,s^{-1}) \geq 43.0$ population and the values from the literature decreases from $\sim0.005$ at $z\sim2.5$ to $\sim0.05$ at $z\sim5-6$. We extrapolate the ratio between the stellar mass densities of $\log_{10} (\rm L_{Ly\alpha}/erg\,s^{-1}) \geq 43.0$ LAEs and LBGs and determine it tends to unity at $z=10$. This implies that these bright LAEs, contribute very significantly to the total stellar mass density during the epoch of reionisation, highlighting the importance of LAEs to the evolution of primeval galaxies in the early Universe.

\section{Conclusions} \label{sec:conclusions}

In this work, we determine the UV luminosity functions (LFs) and stellar mass functions (SMFs) of $\sim4000$ LAEs from the SC4K sample at $z\sim2-6$. Our main results are:

\begin{itemize}

\item M$_{\rm UV}$ and L$_{\rm Ly\alpha}$ are typically correlated (M$_{\rm UV}=-1.6_{-0.3}^{+0.2}\log_{10} (\rm L_{Ly\alpha}/erg\,s^{-1})+47_{-11}^{+12}$) in our sample of LAEs. The relation between M$_\star$ and L$_{\rm Ly\alpha}$ is shallower ($\log_{10} (\rm M_\star/M_\odot)=0.9_{-0.1}^{+0.1}\log_{10} (\rm L_{Ly\alpha}/erg\,s^{-1})-28_{-3.8}^{+4.0}$).

\item Different L$_{\rm Ly\alpha}$ limits significantly affect the shape and normalisation of the UV LF and SMF of LAEs. An increasing L$_{\rm Ly\alpha}$ cut predominantly reduces the number density of lower stellar masses and faint UV luminosities, more significantly for the UV LF. We estimate a proxy for the full UV LF and SMF of LAEs, making simple assumptions of fitting range and faint end slope.

\item For the UV LF of LAEs, we find a characteristic number density ($\Phi^*$) decrease from $\log_{10}(\Phi^*/$Mpc$^{-3})\sim-3.5$ at $z\sim2-4$ to $\sim-4.5$ at $z\sim5-6$, and a brightening of characteristic UV luminosity (M$_{\rm UV}^*$) from -20.6 to -21.8 at the same redshift ranges.

\item For the SMF of LAEs, we measure a decline of $\Phi^*$ with increasing redshift, from $\log_{10}(\Phi^*/$Mpc$^{-3})=-4.6$ at $z=2.5$ to -5.8 at $z=5.4$, and a characteristic stellar mass (${\rm M_\star^*/M_\odot}$) increase from $\log_{10}(\rm M_\star^*/M_\odot)=10.6$ to 11.5 at the same redshift ranges.

\item We apply a uniform luminosity cut of $\log_{10} (\rm L_{Ly\alpha}/erg\,s^{-1}) \geq 43.0$ to our entire sample, producing a subsample of rare bright primeval galaxies. We find a more moderate to no evolution of the UV LF and SMF of this subsample, indicating that the trends computed for the full samples may be driven by differences in the luminosity cuts. This highlights the importance of obtaining deep high-$z$ studies with e.g. MUSE.

\item We compute $\Phi_{\rm LAE}/\Phi_{\rm LBG}$ (proxy of $\chi_{\rm Ly\alpha}$) which tends to unity with increasing M$_{\rm UV}$ at all redshifts, as bright LAEs are typically also bright in M$_{\rm UV}$. For fainter LAEs, the ratio tends to one as we reach fainter Ly$\alpha$ fluxes, with a simple extrapolation implying that by reaching $\log_{10} (\rm L_{Ly\alpha}/erg\,s^{-1})=41.0$ we would approach unit for M$_{\rm UV}=-20$ galaxies at $z=2.5$.

\item The luminosity density ($\rho_{\rm UV}$) shows moderate evolution from 10$^{25.2}$\,erg s$^{-1}$ Hz$^{-1}$ Mpc$^{-3}$ at $z=2.5$ to 10$^{25.0}$\,erg s$^{-1}$ Hz$^{-1}$ Mpc$^{-3}$ at $z\sim5-6$, and the stellar mass density ($\rho_{\rm M}$) decreases from $\sim10^{6.1}$ to $\sim10^{5.5}$\,M$_\odot$\,Mpc$^3$ at the same redshifts. Both $\rho_{\rm UV}$ and $\rho_{\rm M}$ are found to always be lower than the total luminosity and stellar densities of continuum-selected galaxies but slowly approaching it with increasing redshift. Overall, we find our measurements reveal a $\rho_{\rm UV}$ and $\rho_{\rm M}$ of LAEs that slowly approach the measurements of continuum-selected galaxies at $z>6$, pointing to the very significant role of LAEs in the epoch of reionisation.
\end{itemize}

\section*{Acknowledgements}

We thank the anonymous referee for the very constructive feedback which significantly improved the quality and clarity of this work. SS acknowledges a studentship from Lancaster University. Based on data products from observations made with ESO Telescopes at the La Silla Paranal Observatory under ESO programme ID 179.A-2005 and on data products produced by CALET and the Cambridge Astronomy Survey Unit on behalf of the UltraVISTA consortium.

Finally, the authors acknowledge the unique value of the publicly available analysis software {\sc TOPCAT} \citep{Taylor2005} and publicly available programming language {\sc Python}, including the {\sc numpy}, {\sc pyfits}, {\sc matplotlib}, {\sc scipy} and {\sc astropy} \citep{Astropy2013} packages.

\section*{Data availability}

The data underlying this article is based on the public SC4K sample of LAEs \citep[][]{Sobral2018}, available at \href{https://dx.doi.org/10.1093/mnras/sty378}{https://dx.doi.org/10.1093/mnras/sty378}. The derived properties of SC4K LAEs \citep{Santos2020} are available at \href{https://dx.doi.org/10.1093/mnras/staa093}{https://dx.doi.org/10.1093/mnras/staa093}. Additional data presented in this article will be shared on request to the corresponding author.

\bibliographystyle{mnras}
\bibliography{myBib}


\appendix


\bsp	
\label{lastpage}
\end{document}